\documentclass[10pt,journal]{IEEEtran}
\usepackage{cite}
\usepackage{color}

\ifCLASSINFOpdf
  \usepackage[pdftex]{graphicx}
\else
\fi

\usepackage{verbatim}

\usepackage{tabularx}
\usepackage{multirow}
\usepackage{tabularx}
\usepackage[hyphens]{url}
\usepackage{booktabs,floatrow}
\begin{document}
\pagestyle{plain}
\setcounter{page}{1}
\pagenumbering{arabic}
%
\title{Cyber-Physical Systems Security -- A Survey}

%

\author{Abdulmalik Humayed,
        Jingqiang~Lin,
        Fengjun~Li,
        and Bo~Luo 
\thanks{A. Humayed, F. Li and B. Luo were with the Department
of Electrical Engineering and Computer Science, The University of Kansas, Lawrence, KS, 66045 USA.
e-mail: bluo@ku.edu.}
\thanks{Jingqiang Lin was with The State Key Laboratory of Information Security, Institute
of Information Engineering, Chinese Academy of Sciences.}
\thanks{This manuscript is under submission for journal publication.}
}

\maketitle

\begin{abstract}
	With the exponential growth of cyber-physical systems (CPS), new security challenges have emerged. Various vulnerabilities, threats, attacks, and controls have been introduced for the new generation of CPS. However, there lack a systematic study of CPS security issues. In particular, the heterogeneity of CPS components and the diversity of CPS systems have made it very difficult to study the problem with one generalized model.
	
	In this paper, we capture and systematize existing research on CPS security under a unified framework. The framework consists of three orthogonal coordinates: (1) from the \emph{security} perspective, we follow the well-known taxonomy of threats, vulnerabilities, attacks and controls; (2)from the \emph{CPS components} perspective, we focus on cyber, physical, and cyber-physical components; and (3) from the \emph{CPS systems} perspective, we explore general CPS features as well as representative systems (e.g., smart grids, medical CPS and smart cars). The model can be both abstract to show general interactions of a CPS application and specific to capture any details when needed. By doing so, we aim to build a model that is abstract enough to be applicable to various heterogeneous CPS applications; and to gain a modular view of the tightly coupled CPS components. Such abstract decoupling makes it possible to gain a systematic understanding of CPS security, and to highlight the potential sources of attacks and ways of protection.
	
\end{abstract}


\section{Introduction}
\IEEEPARstart{I}{n} recent years, we have witnessed an exponential growth in the development and deployment of various types of Cyber-Physical Systems (CPS). They have brought impacts to almost all aspects of our daily life, for instance, in  electrical power grids, oil and natural gas distribution, transportation systems, health-care devices, household appliances, and many more. Many of such systems are deployed in the critical infrastructure, life support devices, or are essential to our daily lives. Therefore, they are expected to be free of vulnerabilities and immune to all types of attacks, which, unfortunately, is practically impossible for all real-world systems.

One fundamental issue in CPS security is the heterogeneity of the building blocks. CPS are composed of various components in many ways. There are different hardware components such as sensors, actuators, and embedded systems. There are also different collections of software products, proprietary and commercial, for control and monitoring. As a result, every component, as well as their integration, can be a contributing factor to a CPS attack. Understanding the current CPS security vulnerabilities, attacks and protection mechanisms will provide us with a better understanding of the security posture of CPS. Consequently, we should be able point out the limitations of CPS that make them subject to different attacks and devise approaches to defend against them.

The complexity of cyber physical systems and the heterogeneity of CPS components have introduced significant difficulties to security and privacy protection of CPS. In particular, with the complex \emph{cyber-physical interactions}, threats and vulnerabilities becomes difficult to asses, and new security issues arise. It is also difficult to identify, trace and examine the attacks, which may originate from, move between, and target at multiple CPS components. An in-depth understanding of the vulnerabilities, threats and attacks is essential to the development of defense mechanisms. A survey of existing CPS security and privacy controls will also enable us to identify missing pieces, weak links and new explorations.

\begin{figure}[ht!]
\includegraphics[width=90mm]{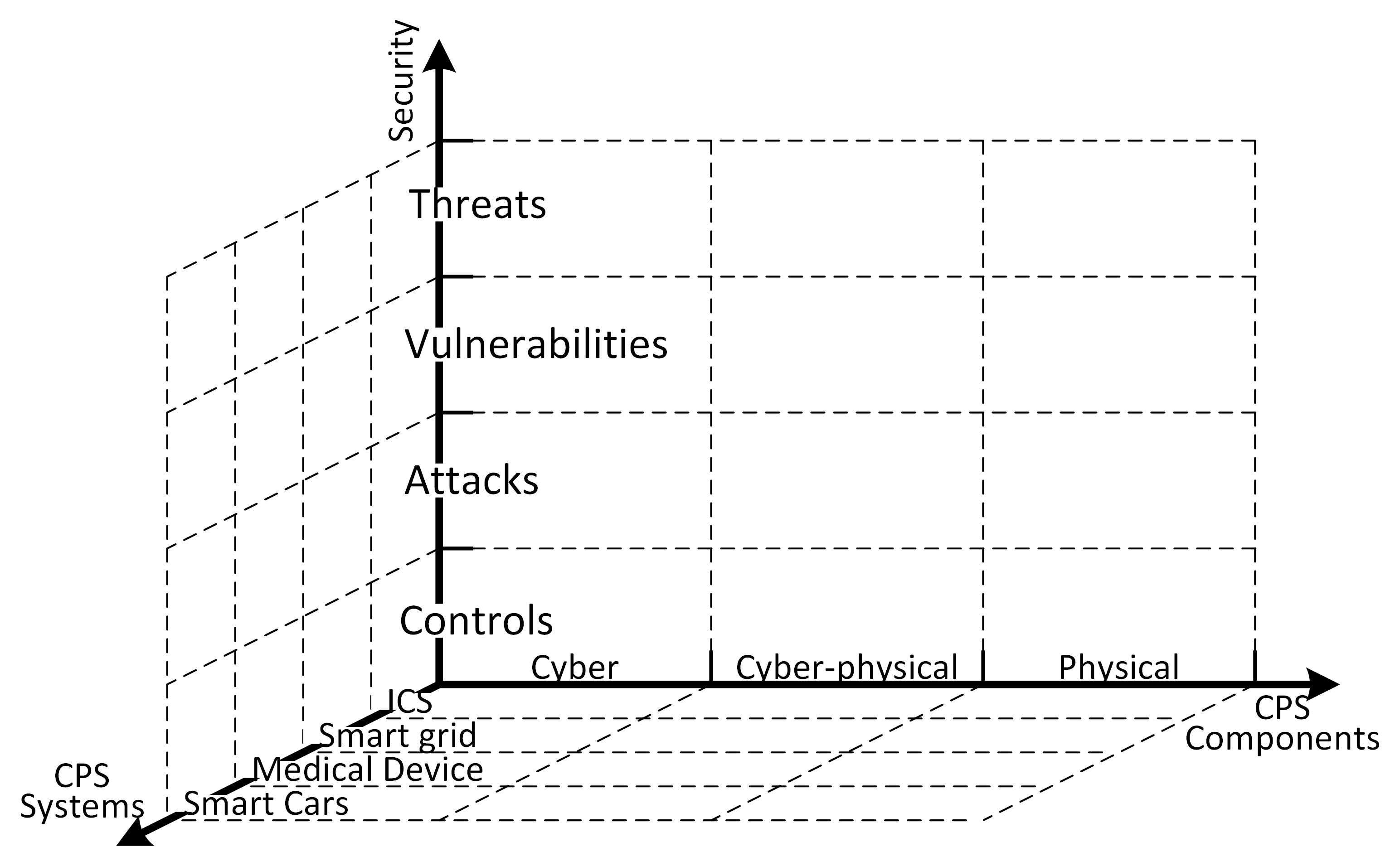}
\caption{CPS security framework with three orthogonal coordinates: security, CPS components, and representative CPS systems.}	
\label{cordin}
\end{figure}

In this survey, we first briefly introduce CPS, with a special focus on how they are different from either legacy control systems or traditional IT systems. Recognizing the difference is key in understanding CPS security problems. We then survey the literature on CPS privacy and security under a unified framework, which consists of three orthogonal coordinates, as shown in Figure \ref{cordin}. First, from \emph{security} perspective, we follow the well-known taxonomy of \emph{threats} (Section \ref{sec:threats}), \emph{vulnerabilities} (Section \ref{vulns}), \emph{attacks} (Section \ref{attacks}) and \emph{controls} (Section~\ref{sec:controls}). Next, we discuss each main aspect following the \emph{CPS components} perspective: \emph{cyber}, \emph{physical}, and \emph{cyber-physical}. For instance, when we survey the attacks, we categorize them into cyber-attacks, physical-attacks, and cyber-physical-attacks. Last, from the \emph{CPS systems} perspective, we explore general CPS features as well as representative systems, in particular, industrial control systems, smart grids, medical CPS, and smart cars. At the end of Section~\ref{sec:controls}, we summarize the key threats, vulnerabilities, attacks and controls in each CPS aspect for each representative CPS system. In this survey, we not only systematizes existing knowledge and provide insightful perspectives on CPS security, but also identify open areas that need more attention, and highlight the unanswered challenges (Section \ref{sec:challenges}).

In this work, our contributions are as follows: (1) We propose a CPS security framework that aims to distinguish between cyber, cyber-physical, and physical components in a given system. (2) We survey potential threat sources and their motivations. (3) We present the existing vulnerabilities and highlight the root reasons with actual examples. (4) We survey reported attacks on CPS and pinpoint the underlying vulnerabilities and subtly influenced CPS components. (5) We also summarize existing control mechanisms, and further identify the unsolved issues and challenges in different CPS applications.

\section{Background}
\label{overview}
\subsection{Cyber-Physical Systems}
While there doesn't exist a unanimously accepted, authoritative definition of Cyber Physical Systems (CPS), we can simply say that CPS are systems used to monitor and control the physical world. They are perceived as the new generation of embedded control systems such that CPS are networked embedded systems. In addition, systems, where sensor and actuator networks are embedded, are also considered CPS~\cite{Cardenas2011Attacks}. Because of the reliance on IT systems, CPS could be defined as IT systems that are integrated into physical world application~\cite{Gollmann2013Security}. This integration is a result of the advancements in the information and communication technologies (ICT) to enhance interactions with physical processes. All of these definitions highlight the heavy presence of the interactions between the cyber and the physical worlds.


An increasing dependence on CPS is growing in various applications such as energy, transportation, military, health-care, and manufacturing. CPS can be called different names, depending on the application using them. For example, a very important and representative CPS is the Supervisory Control and Data Acquisition (SCADA) system, which is used in Critical Infrastructure (CI) such as the Smart Grid and Industrial Control Systems (ICS). Other examples have emerged in medical devices such as wearable and implantable medical devices. In addition, a network of small control systems are embedded in modern cars to improve fuel efficiency, safety, and convenience. Here we introduce briefly four representative applications of CPS that we will cover throughout the paper.

\vspace{2mm}\noindent\textbf{Industrial Control Systems (ICS)}. ICS refers to control systems used to enhance the control, monitoring, and production in different industries such as the nuclear plants, water and sewage systems, and irrigation systems. Sometimes ICS is called Supervisory Control and Data Acquisition (SCADA) or Distributed Control Systems (DCS). For consistency, we will use the term ICS hereafter. In ICS, different controllers with different capabilities collaborate to achieve numerous expected goals. A popular controller is the Programmable Logic Controller (PLC), which is a microprocessor designed to operate continuously in hostile environments~\cite{lee2011introduction}. This field device is connected to the physical world through sensors and actuators. Usually, it is equipped with wireless and wired communication capacity that is configured depending on the surrounding environments. It can also be connected to PC systems in a control center that monitors and controls the operations.

\vspace{1mm}\noindent\textbf{Smart Grid Systems}.
The smart grid is envisioned as the next generation of the power grid that has been used for decades for electricity generation, transmission, and distribution. The smart grid provides several benefits and advanced functionalities. At the national level, it provides enhanced emission control, global load balancing, smart generation, and energy savings. Whereas at the local level, it allows home consumers better control over their energy use that would be beneficial economically and environmentally~\cite{McDaniel2009SPIEEE}. The smart grid is comprised of two major components: power application and supporting infrastructure~\cite{Sridhar2012IEEE}. The power application is where the core functions of the smart grid are provided, i.e., electricity generation, transmission, and distribution. Whereas the supporting infrastructure is the intelligent component that is mainly concerned with controlling and monitoring the core operations of the smart grid using a set of software, hardware, and communication networks.

\vspace{1mm}\noindent\textbf{Medical Devices}.
Medical devices have been improved by integrating cyber and physical capabilities to deliver better health care services. We are more interested in medical devices with cyber capabilities that have physical impact on patients. Such devices are either implanted inside the patient's body,  called Implantable Medical Devices (IMDs), or worn by patients, called wearable devices. They are usually equipped with wireless capabilities to allow communication with other devices such as the \textit{programmer}, which is needed for updating and reconfiguring the devices. Wearable devices communicate with each other or with other devices, such as a remote physician or smartphone~\cite{Rushanansok2014SoK}.

\vspace{1mm}\noindent\textbf{Smart Cars}.
Smart cars (intelligent cars) are cars that are more environment-friendly, fuel-efficient, safe, and have enhanced entertainment and convenience features. These advancements are made possible by the reliance on a range of 50 to 70 computers networked together, called Electronic Control Units (ECUs). ECUs are responsible for monitoring and controlling various functions such as engine emission control, brake control, entertainment (radio, multimedia players) and comfort features (cruise control and windows opening and closing).

\subsection{CPS Communications}
Communication technologies vary in CPS applications. Different application use different protocols, open and proprietary, and technologies, wired and wireless. Here we give a brief overview of the most common communication technologies and protocols in each of the four applications.

\vspace{1mm}\noindent\textbf{ICS}.
Two categories of communication protocols are deployed in ICS, one is used for the automation and control such as Modbus, Distributed Network Protocol (DNP3), and the other is for interconnecting ICS control centers, such as Inter-Control Center Protocol (ICCP)~\cite{alcaraz2013critical}. Those protocols are used in addition to general-purpose protocols such as TCP/IP.

\vspace{1mm}\noindent\textbf{Smart Grid}.
The networks are of two types: field device communications within substations using Modbus and DNP3, and recently the more advanced protocol, developed by the International Electrotechnical Commission (IEC), IEC 61850. The other type is control center communications, which also rely on ICCP, similar to ICS. In addition, smart meters and field devices use wireless communications to send measurements and receive commands from control centers. Smart meters, for example, use short-range frequency signals, e.g., Zigbee, for diagnostics operations by technicians or readings by digital smart readers.

\vspace{1mm}\noindent\textbf{Medical Devices}.
It is a necessary requirement that IMDs be configured and updated wirelessly, so that no surgical extraction for the device is needed. Therefore, wireless communication is the most common method of communication in medical devices. IMDs and wearable devices rely on different communication protocols and technologies. For example, IMDs use low frequency (LF) signals specified by The Federal Communications Commission (FCC),  called Medical Implant Communication Service (MICS), that make it possible for IMDs and their programmers to communicate. On the other hand, wearable devices rely on another type of wireless communications, i.e., Body Area Network (BAN). BAN utilizes a number of wireless communication technologies such as Bluetooth and ZigBee~\cite{Chen2011Body}.

\vspace{1mm}\noindent\textbf{Smart Cars}.
Smart cars can have different types of communication capacities, including Vehicle to Vehicle (V2V), Vehicle to Infrastructure (V2I), and in-vehicle communications. In this paper we focus on the latter. As we mentioned, cars have around 70 connected ECUs, all of which communicate through a bus network. The network is usually divided into multiple subnetworks, each of which also has a bus topology. Subnetworks can exchange messages through a gateway that separates their traffics. A common conception is that this separation is due to security concerns. However, ~\cite{Checkoway2011Comprehensive} suggest that this is also for bandwidth concerns. The most common protocols are (1) the Local Interconnect Network (LIN), used for relatively low speed applications such as opening/closing windows; (2) Controller Area Network (CAN), used for soft real-time applications such as the anti-lock braking system; (3) Flexray, needed for hard real-time applications where the speed of transmission is critical such as braking or responding to an obstacle in front of the car; and (4) Media Oriented Systems Transport (MOST), used for in-car entertainment applications~\cite{Wolf2004Securityin}. In addition, some cars are equipped with wireless connections such as Bluetooth and cellular interfaces.

\subsection{CPS Models and Aspects}
\label{CPSComps}

\begin{figure}[h]
 	\includegraphics[width=80mm]{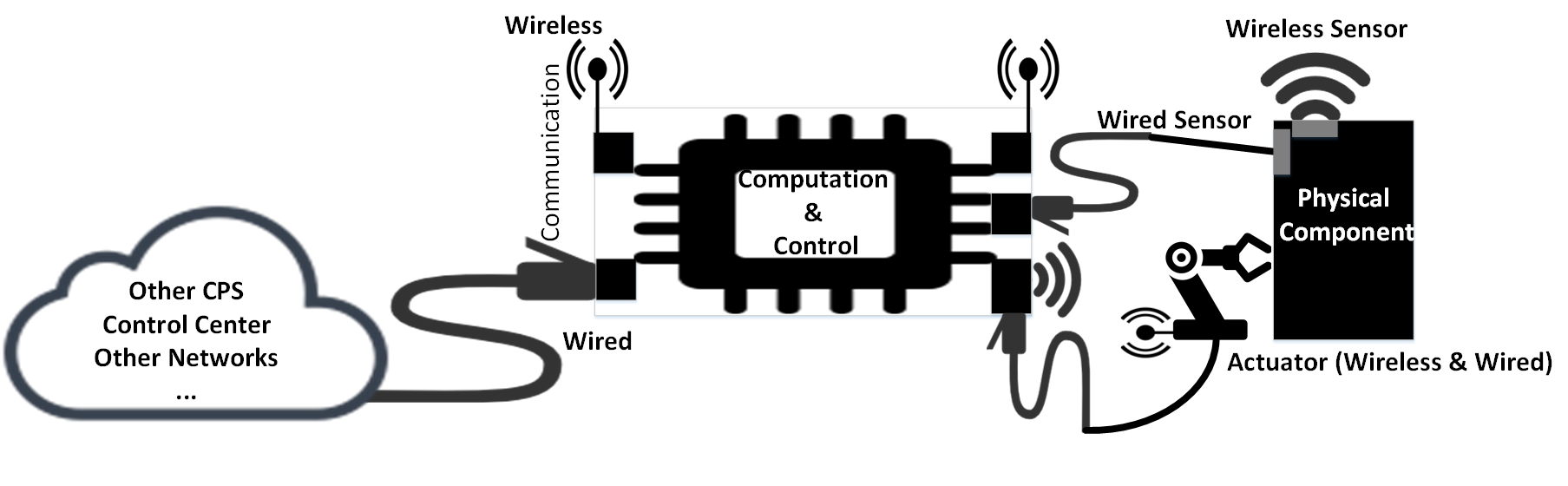}
 	\caption{CPS abstract model}
 	\label{CPSModel}
 \end{figure}

Fig.~\ref{CPSModel} shows a high-level abstraction of any Cyber Physical System, which mainly consists of three categories of components: (1) communication, (2) computation and control, and (3) monitoring and manipulation. The communication could be wireless or wired, and it could connect CPS with higher-level systems, such as control centers, or with lower-level components in the physical world. The computation and control part is where the intelligence is embedded, control commands are sent, and sensed measures are received. The monitoring and manipulation components connect CPS to the physical world through sensors to monitor physical components, and actuators to manipulate them.

A CPS component might have the ability to communicate with control centers or other CPS components. This same component could also contain a sensor, an actuator, or both to connect to the physical world. Each one of these capabilities has different security implications that may result from the interactions of the component's parts and their capabilities. For example, a CPS component's communication and computational functions are not expected to affect the physical world, and yet might be exploited to cause unexpected attacks with physical consequences. Similarly, the physical properties of this component, in addition to the physical properties of the object of interest in the physical world that the CPS controls and monitors, can also cause unexpected attacks that might result in non-physical attacks, such as misleading information sent to a CPS.

This heterogeneity of CPS, among components, or within a component itself, results in a lack of understanding of new types of security threats that would exploit such heterogeneity. The need to clearly distinguish between such aspects for security analysis and engineering arises. Thus, we propose to view any CPS from three aspects: \emph{cyber}, \emph{cyber-physical}, and \emph{physical}. The cyber aspect considers data computations, communications, and interactions that do not affect the physical world, whereas the cyber-physical aspect considers all interactions with the physical world. The cyber-physical aspect is where the cyber and physical world can connect. Finally, the physical aspect includes any physical components that their properties might have security-related exploitations.

\begin{figure}[h]
 	\includegraphics[width=85mm]{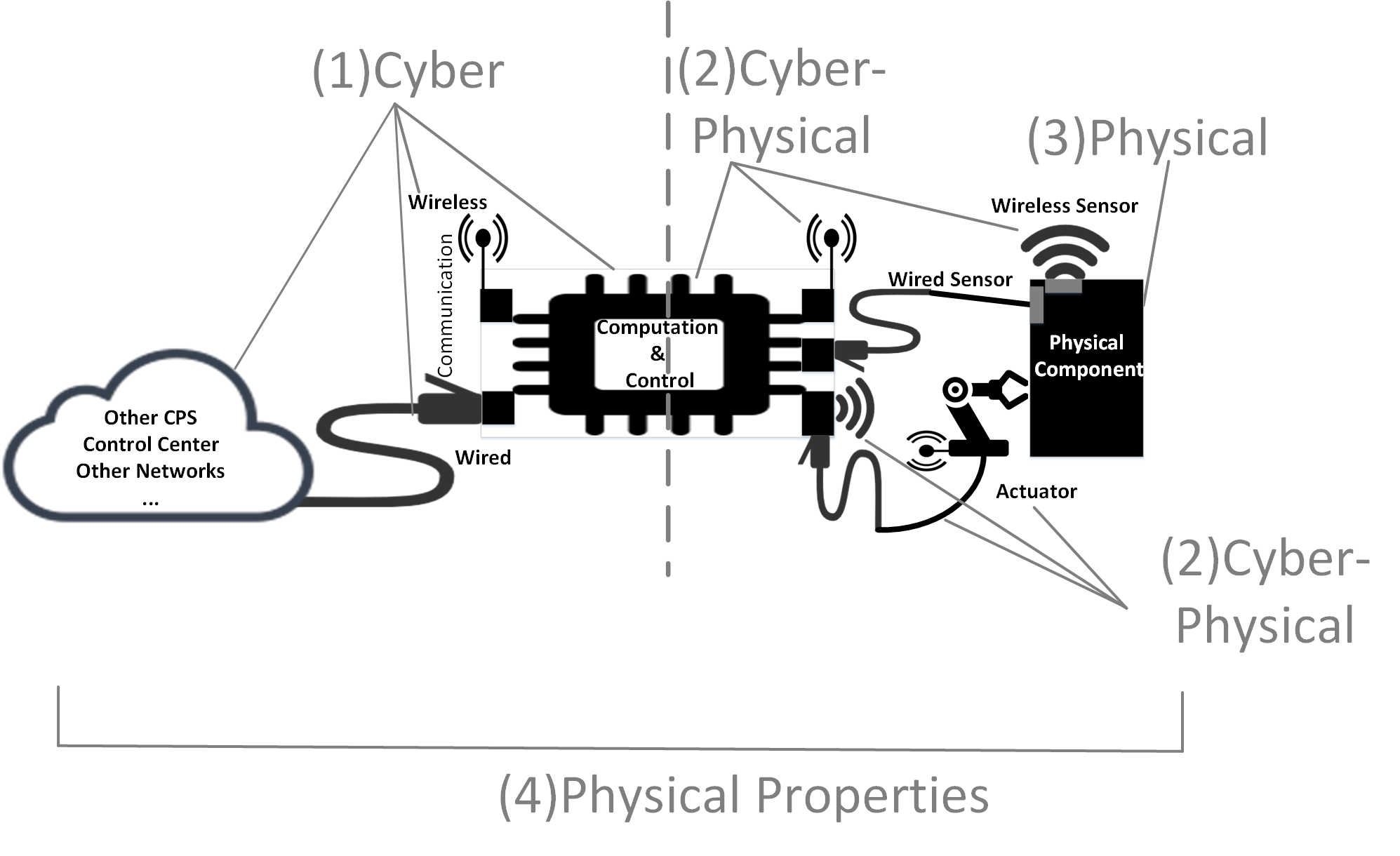}
 	\caption{CPS Aspects}
 	\label{CPSAspects}
 \end{figure}

In Figure~\ref{CPSAspects}, we incorporated the aforementioned CPS view in the annotated figure shown in Fig~\ref{CPSModel}. In Figure~\ref{CPSAspects}, (1) indicates aspects that we consider cyber, whereas (2) denotes cyber-physical aspects. Note the dashed line separating (1) and (2) shows how the same component can be considered cyber and cyber-physical at the same time depending on the presence or absence of the interaction with the physical world. (4) shows that the physical properties of any part of a CPS system could play a role in security issues. Therefore, we need to include them in the physical aspect.

In the following paragraphs, we present how our abstract model can capture the CPS aspects in the representative applications. For each application, we show a figure annotated it with the CPS aspects: (1) cyber, (2) cyber-physical, and (3) physical. 

\begin{figure}[ht!]
\includegraphics[width=80mm]{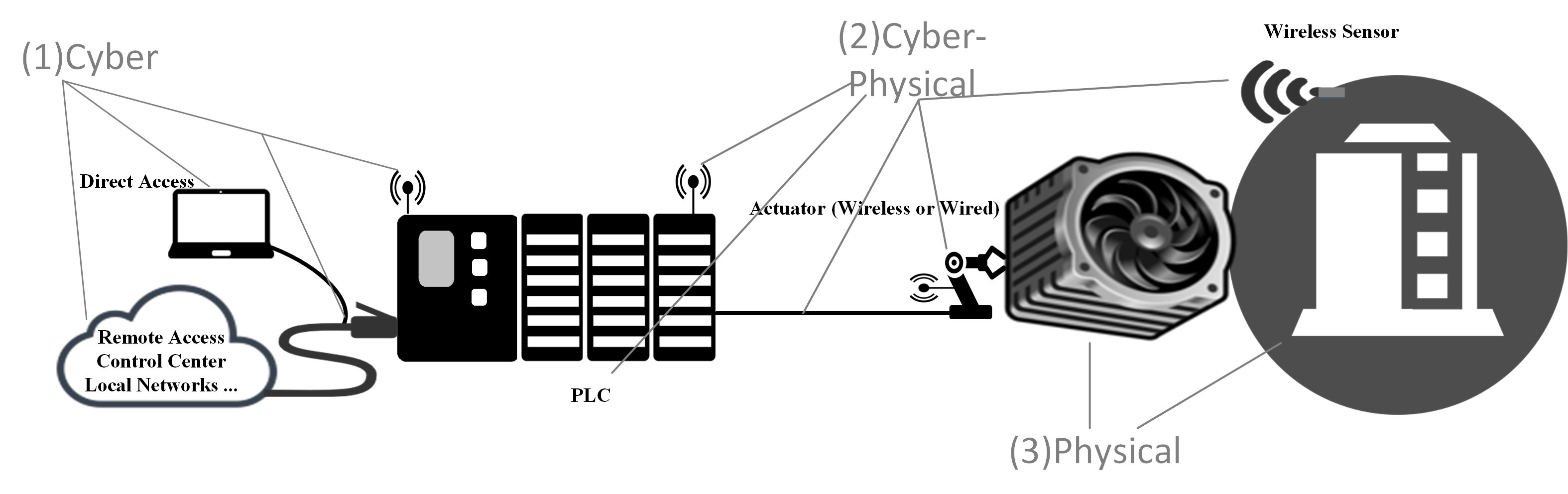}
\caption{CPS aspects in ICS}
\label{PLC}
\end{figure}

\vspace{1mm}\noindent\textbf{ICS}.
Figure~\ref{PLC} depicts the CPS aspects in a PLC scenario, where it is used for controlling the temperature in a chemical plant. The goal is to maintain the temperature within a certain range. If the temperature exceeds a specified threshold, the PLC is notified via a wireless sensor attached to the tank, which in turn, notifies the control center of the undesired temperature change. Alternatively, in closed-loop settings, the PLC could turn the cooling system on to reduce that tank's temperature within the desired range.

In this figure, the cyber aspects (1) are the cyber interactions with the PLC such that there is no direct interaction with physical components, such as cooling fans or the tank. This involves laptops that can directly connect PLCs, communications with higher-level environments such as the control center and other remote entities, and the PLC's wireless interface that could be based on long- or short-range frequencies. In addition, cyber-physical aspects (2) are those that connect cyber and physical aspects. The PLC, the actuator, and the sensor, are all cyber-physical aspects due to their direct interactions with the physical world. The wireless capabilities of the actuator and the sensor are also considered cyber-physical. Finally, the physical aspects are the physical objects that need monitoring and control, i.e. the cooling fans and the tank's temperature.

\begin{figure}[ht!]
	\includegraphics[width=80mm]{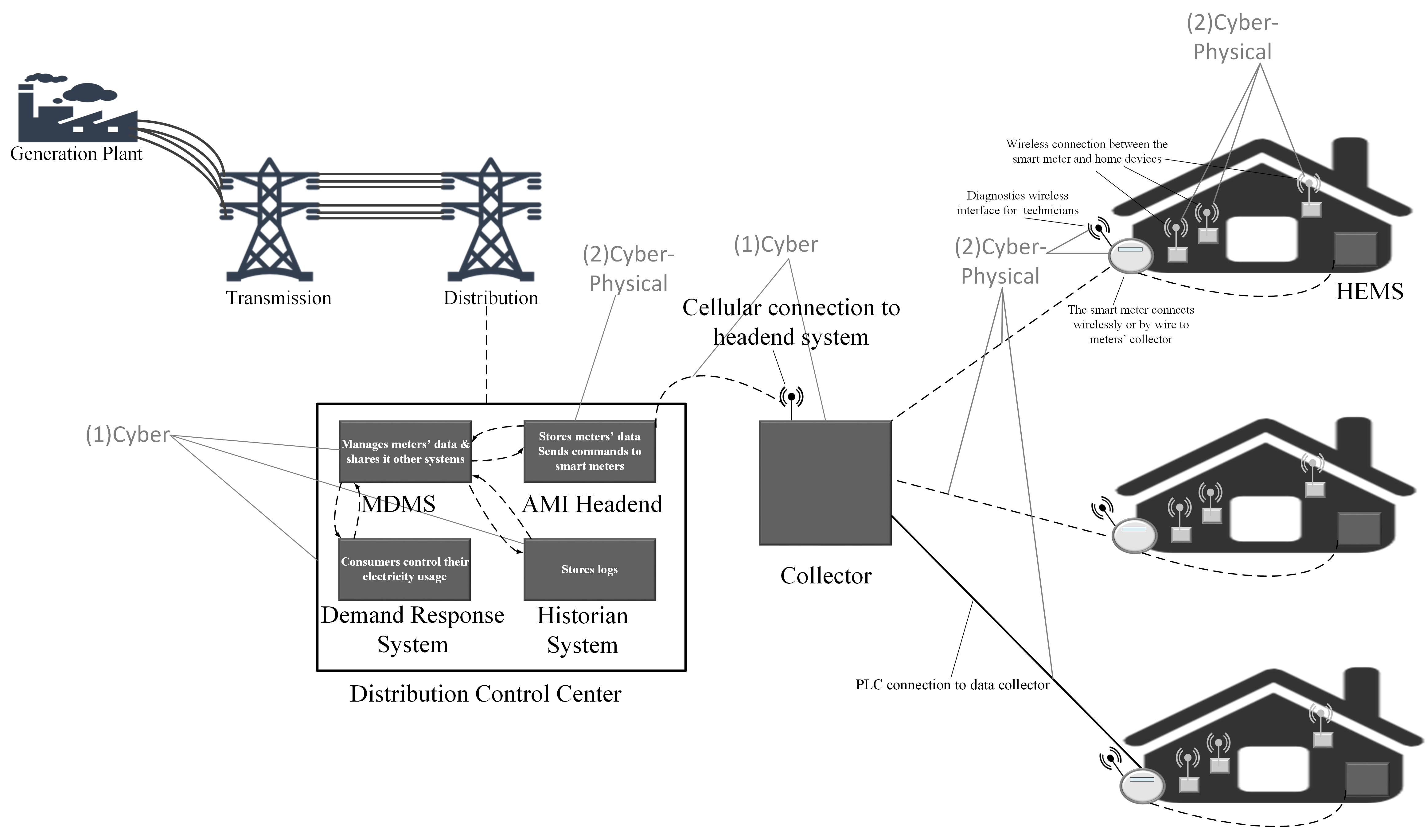}
	\caption{CPS aspects in the Smart Grid}
	\label{SmartMeter}
\end{figure}

\vspace{1mm}\noindent\textbf{Smart Grid}.
Figure~\ref{SmartMeter} shows a typical scenario in smart grids. A smart meter is attached to every house to provide utility companies with more accurate electricity consumption data and customers with convenient way to track their usage information. A smart meter interfaces a house's appliances and Home Energy Management Systems (HEMS) on the one hand, and interfaces with data collectors on the other. Wireless communications are the most common means to communicate with collectors, although wired communications, such as Power Line Communications (PLC), are also available. A meter is equipped with a diagnostics port that relies on short-range wireless interface for convenient access by digital meter readers and diagnostics tools~\cite{Knapp2013Applied}. The smart meter sends the measurements to a collector that aggregates all meters' data in a designated neighborhood. The collector sends the aggregated data to a distribution control center managed by the utility company. In particular, to the AMI headend server that stores the meters' data and shares the stored data with the Meter Data Management System (MDMS) that manages the data with other systems such as demand response systems, historians, and billing systems. The headend can connect/disconnect services by remotely sending commands to the meters. This feature is a double-edged sword such that it is very efficient way to control services, yet it could be exploited to launch large-scale blackouts by remotely controlling a large number of smart meters.

In Fig.~\ref{SmartMeter}, we highlight the CPS aspects in the involved components that have some interactions with the smart meters. Cyber aspects (1) appear in the control center where smart meters' data is stored, shared, and analyzed and based on that some decisions can be made based on the analysis. The control center can also have a cyber-physical aspect (2) when connect/disconnect commands are sent by the AMI headend to smart meters. In addition, the cyber-physical aspect (2) is also apparent in the smart meter itself due to its ability to perform cyber operations, such as sending measurements to utility, and physical operations, such as connecting/disconnecting electricity services. Other field devices in the generation, transmission automation, and distribution plants have a high presence of the cyber-physical aspect due to their close interactions with physical aspects of smart grids. Home appliances that are connected with smart meters are considered cyber-physical because of the their direct interaction with smart meters. A utility company can use smart meters to
control the amount of energy consumed by particular home appliances when needed~\cite{Neuman2009Challenges}, which is a cyber-physical (2) action.

%
%
%

\begin{figure}[ht!]
 	\includegraphics[width=70mm]{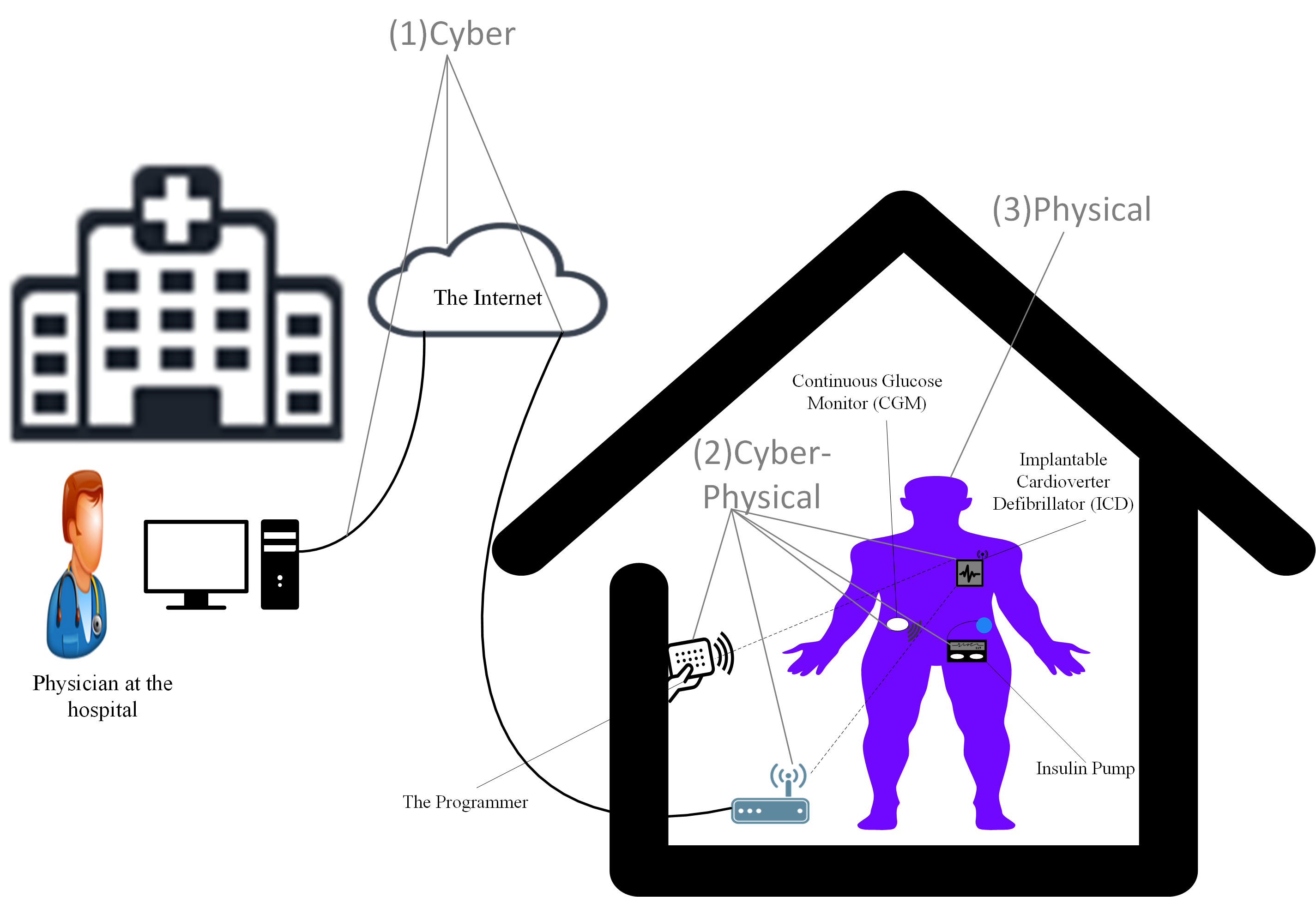}
 	\caption{CPS aspects in medical devices}	
 	\label{MedicalFig}
 \end{figure}

\vspace{1mm}\noindent\textbf{Medical Devices}. Fig.~\ref{MedicalFig} is an overview of two of the most popular IMDs, the \textit{insulin pump} and the \textit{implantable cardioverter defibrillator} (ICD). The insulin pump is used to automatically or manually inject insulin injections for diabetics when needed, whereas the ICS is used to detect rapid heartbeat and response by delivering an electric shock to maintain a normal heartbeat rate~\cite{Halperin2008Pacemakers}. The insulin pump usually needs another device, called the \textit{continuous glucose monitor} (CGM), to receive blood sugar measurements. Both devices, the insulin pump and the CGM, require small syringes to be injected into a patient's body. The insulin pump receives measurements of glucose levels from the CGM. Based on the measurements, the pump decides whether the patient needs an insulin dose or not. The CGM sends the measurements through wireless signals to the insulin pump or other devices, such as a remote control or computer. In addition, some insulin pumps can be commanded by a remote control held by a patient or physician.

In this figure, the cyber aspects (1) are embodied in the monitoring computers in the hospital and the communications to the Internet. The cyber-physical aspects (2), on the other hand, are present in those devices that directly interact with patients' implanted devices. A patient represents the physical aspect (3) in the context of medical devices. An IMD connects to the hospital by sending measurements through an in-home router. In order to reconfigure an ICD, a physical proximity is required to be able to do so using a device called the \textit{programmer}.

\begin{figure}[ht!]
\includegraphics[width=60mm]{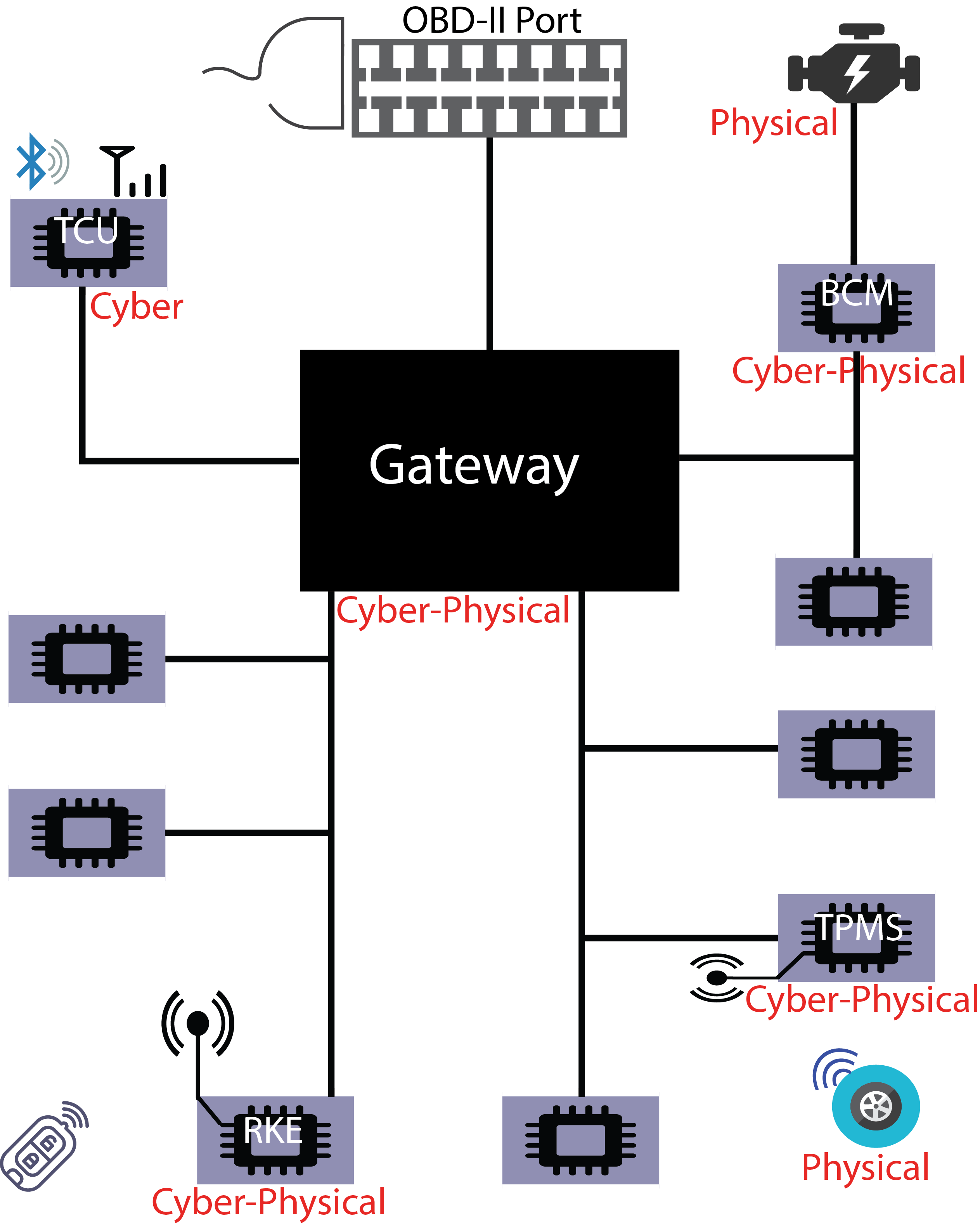}
\caption{CPS aspects in smart cars}	
\label{ECU}
\end{figure}

\vspace{1mm}\noindent\textbf{Smart Cars}.
Figure~\ref{ECU} shows the typical architecture of an in-car network. Depending on the nature of the tasks expected from each ECU, an ECU is attached to the appropriate subnetwork. ECUs from different subnetworks can intercommunicate through gateways. In this paper we mainly focus on CAN bus for two reasons: 1) most security issues result from CAN-based networks and 2) it has been required to be deployed in all cars in the U.S. since 2008~\cite{Experimental2010Koscher}, thus it is in almost every car around us.

In Fig.~\ref{ECU}, we annotated ECUs that do not have any interactions with physical components of the cars as cyber (1). Examples of which include the Telematics Control Unit (TCU) and the media player. The TCU has more than a wireless interface that allows advanced capabilities such as remote software updates by car manufacturers, phone pairing, hands-free usage of phones. The cyber-physical (2) annotations are for ECUs that can legitimately interact with physical components and manipulate them, such as the parking assist and the Remote Keyless Entry (RKE) systems. The RKE, for example, receives signals to make a physical impact on the car by locking/unlocking doors. Finally, physical components such as the engine or tires are physical (3).

\subsection{Security in CPS}

In this section, we motivate the importance of security in CPS with four specific illustrative examples. Security control is usually associated with mechanisms such as cryptography, access control, intrusion detection, and many other solutions commonly used in IT systems. Those mechanisms are very important in securing information and communication technology's infrastructure. However, many reported attacks on CPS applications show the inadequacy of the sole dependence on these mechanisms as presented in Section~\ref{attacks}. Therefore, solutions that take cyber-physical aspects into account are needed and could be complemented with IT security solutions.


\vspace{1mm}\noindent\textbf{Security in ICS}.
Lack or weakness of security in CPS could be catastrophic depending on the application. For example, if the security of CPS used in a nuclear plant has been compromised, a world-wide threat is the possible consequence. Furthermore, security violations in smart grids could lead to the loss of services to the consumer and financial losses to the utility company. Because of the CPS's pervasiveness and its wide use in the critical infrastructure in particular, CPS security is of a critical importance. In fact, it is even suggested that ICS is not yet ready to be connected to the Internet~\cite{FranciaIII2012Security}. This is due to the inherent security vulnerabilities in the legacy control systems and their communications.

\vspace{1mm}\noindent\textbf{Security in Smart Grids}.
Adequate security in smart grids poses the threat of remote attacks that could result in large-scale blackouts. Blackouts could result in safety implications such as medical equipments malfunctions, loss of data in data centers, and even an increase in crime rate~\cite{Eaton2014Power}. Another security inadequacy could result in compromised privacy such as attackers' ability to reveal customers' personal information.

\vspace{1mm}\noindent\textbf{Security in Medical Devices}.
Security in wearable and IMDs makes them immune to attacks that might compromise patients' safety and privacy.Because of the different circumstances surrounding medical devices, the need for defining appropriate security goals arises. Halperin et al. ~\cite{Halperin2008Security} initiated the discussion of the security goals in medical devices by extending the standard security goals, confidentiality, integrity, and availability. Security goals include the authorized entities should be able to access accurate data, identify and configure devices, update software, and maintain the device's availability; whereas privacy goals include the protection of private information about a device's existence, type, unique ID, and patient's identification.


\vspace{1mm}\noindent\textbf{Security in Cars}.
 Car manufacturers strive to come up with a variety of innovative technologies that would satisfy their customers by providing more functionalities and comfort. Typically cars are safe by design, but security, however, is not usually of a great concern in the design phase. Safety ensures the car's ability to function during non-malicious incidents. Security, on the other hand, has not been a design issue, but rather an add-on feature. The new features in cars require wireless communications and components with physical impacts. These two features alone result in most security vulnerabilities and attacks in smart cars.

\section{CPS Security Threats}\label{sec:threats}

Securing CPS bears with it various challenges, one of which is understanding the potential threats~\cite{Cardenas2009Challenges}. We aim to tackle this challenge by identifying CPS potential threats and shedding light on them from different angles. First we discuss the general threats that almost any CPS application could be vulnerable to. Then we dive into various threats that are more specific to each CPS application. Traditionally, in order for a system to be secure, it satisfies the three security requirements: confidentiality, integrity and availability. Due to the different nature of CPS and their direct interaction with the physical world, safety requirements are also crucial. Here we discuss the threats to both security and safety of CPS.

\subsection{General CPS Threat Model}
The knowledge of who/what we protect a CPS from is equally important to the knowledge of the existing vulnerabilities and attack mechanisms. We first need to define what we mean by a \textit{threat}. A security \textit{threat} is defined as ``a set of circumstances that has the potential to cause loss or harm"~\cite{Pfleeger2006Security}. The potentiality aspect is key in this context, as we discuss potential threats that may not necessarily have occurred, but might. The loss might be in safety measures, confidentiality, integrity, or availability of resources, whereas the harm implies harming people, the environment, or systems. Note that due to the pervasiveness of the CPS applications, people are increasingly becoming a critical asset to protect, in addition to the other informational and communicational assets that are common in security literature.

We identify five factors about every threat: \emph{source}, \emph{target}, \emph{motive}, \emph{attack vector}, and \emph{potential consequences}. Then we elaborate on each one by showing possible types applicable to each factor.

\noindent\textbf{(1) Source}. The source of a threat is the initiator of an attack. Threat sources fall into three types: \textit{adversarial threats} which pose malicious intentions from individuals, groups organizations or states/nations; \textit{accidental threats} are threats that have been caused accidentally or through legitimate CPS components; \textit{environmental threats} which include natural disasters (floods, earthquakes), human-caused disasters (fires, explosions), and failures of supporting infrastructure (power outage or telecommunications loss)~\cite{Nicholson2012ComandSec,Ross2012SP800,CERT_ThreatSource,Stouffer2011NIST82,Kang2009TDCE,Cebula2010Taxonomy,Stoneburner2002SP800}.

\noindent\textbf{(2) Target}. Targets are CPS applications and their components or users. We will see specific examples for each application.

\noindent\textbf{(3) Motive}. CPS attackers usually have one or more reasons to launch an attack: criminal, spying, terroristic, political, or cyberwar~\cite{Setola2011Cyber,CERT_ThreatSource}.

\noindent\textbf{(4) Attack Vector}. A threat might perform one type or more of four mechanisms for a successful attack: interception, interruption, modification or fabrication~\cite{Pfleeger2006Security}

\noindent\textbf{(5) Consequence}. Compromising the CPS's confidentiality, integrity, availability, privacy, or safety.

\subsection{CPS Security Threats}

We explore the potential threats to the four CPS applications using the proposed threat model. In particular, we highlight the threats that are specific to each application with respect to the five factors: source, target, motive, vector, and consequence.

\vspace{1mm}\noindent\textbf{Threats against ICS}

\begin{itemize}
\item\emph{Criminal Attackers (motive)}.
An attacker whose familiar with the system (source) could exploit wireless capabilities (vector) to remotely control an ICS application and possibly disrupt its operations (consequence).
\item\emph{Financially-motivated customers (motive)}. A capable customer (source) aiming to reduce a utility bill might be able to tamper with a physical equipment or inject false data (vector) to misinform the utility (target) causing it to lose financially (consequence)~\cite{turk2005cyber}.

\item\emph{Politically-motivated espionage (motive)}. Intelligence agencies (source) might perform reconnaissance operations targeting a nation's critical infrastructure (target) possibly through spreading malware (vector) resulting in confidentiality violations of critical data (consequence)~\cite{Miller2012SIGITE,Munro2012Deconstructing}.
\item\emph{Politically-motivated cyberwar (motive)}. A hostile nation (source) could initiate a cyberwar against another nation (target) by remotely attacking its critical infrastructure, e.g., nuclear plants and gas pipelines, by spreading malware or accessing field devices (vector) resulting in a plant's shutdown, sabotaging components, or environmental pollution (consequence) ~\cite{Ryu2009Reducing,tsang2010cyberthreats,Byres2004Myths,Kang2009TDCE,Krotofil2013INDIN}.
\item\emph{Physical threats}. An attacker (source) could spoof a sensor that measures the temperature of a particular environment (target) by applying heat or cold to it (vector)  resulting in sending misleading false measurements to the control center (consequence).
\end{itemize}

\vspace{1mm}\noindent\textbf{Threats against Smart Grids}
\begin{itemize}
\item\emph{Financially-motivated threats (motive)}. A customer (source) who wants to trick a utility company's billing system (target) might tamper with smart meters (vector) to reduce the electricity bill (consequence)~\cite{Rahman2012Smart,Metke2010Security,McDaniel2009SPIEEE,Mo2012IEEE,Anderson2010Controls}. Another example of this type of threat is when utility companies (source) might be interested in gathering customers' private information (target) by analyzing their electricity usage to infer habits and types of house appliances (vector) in order to sell such information for advertisement purposes resulting in privacy violation (consequence)~\cite{McDaniel2009SPIEEE,quinn2009privacy,chow2011enhancing,Sridhar2012IEEE}. In addition, there is a possible scenario where criminals (source) extort by demanding a ransom (vector) in exchange for not taking down a number of smart meters (target) that might cause a blackout (consequence)~\cite{Anderson2010Controls}.

\item\emph{Criminally or financially-motivated threat (motive)}. Thieves (source) who aim to rob a house (target) might be able to infer private information, such as a house inhabitant's presence, from the communications between the smart meter and the utility company (vector) in order to perform a successful robbery (consequence)~\cite{Sridhar2012IEEE}.

\item\emph{Political threats (motive)}. A hostile nation (source) might initiate a cyberwar against another country's national power system (target) by gaining remote access to the smart grids' infrastructure (vector) resulting in large scale blackouts, disturbances, or financial losses (consequence)~\cite{McDaniel2009SPIEEE}.
\end{itemize}

\vspace{1mm}\noindent\textbf{Threats against Medical Devices}

\begin{itemize}
\item\emph{Criminal threats (motive)}. A criminal hacker (source) might aim to harm a patient and affect his/her health condition (target) by using wireless tools to inject or retransmit previously-captured legitimate commands (vector) in order to change the device's state and expected operations resulting in an undesired health condition (consequence)~\cite{Halperin2008Security}. In addition, an attacker (source) might also be able to cause harm (target) by jamming the wireless signals exchanged between medical devices to maintain a stable health condition (vector) resulting in the unavailability of the device and its failure to deliver the expected therapies (consequence)~\cite{Halperin2008Security,Hanna2011Take,Rushanansok2014SoK}.

\item\emph{Spying threats (motive)}. A hacker (source) aiming to reveal the existence of a disease, a medical device, or any other information that a patient considers private (target) by intercepting the communications of a patient's medial device via wireless hacking tools (vector), which results in a violation of privacy and confidentiality(consequence)~\cite{Halperin2008Security}. In addition, as the medical devices communicate with other parties, such as hospitals, a large amount of private data is stored in various locations. This could tempt an attacker (source) with spying motivations (motive) to gain an unauthorized access to such data (target) through penetrating the networks that connect among the involved legitimate parties (vector) resulting in privacy invasion (consequence)~\cite{Lee2012Challenges}.

\item\emph{Politically-motivated threats (motive)}. Cyberwar has a new attack surface by which a hostile nation (source) could target political figures (target) by attacking their medical devices exploiting the devices wireless communications (vector) resulting in a potential critical health condition or eventual death (consequence)~\cite{Lisa2013Doctors}. In fact, former U.S. Vice President Dick Cheney had the wireless capabilities disabled in his pacemaker because he was aware of the possible realistic assassination threats~\cite{Rushanansok2014SoK}.
\end{itemize}

\vspace{1mm}\noindent\textbf{Threats against Smart Cars}

\begin{itemize}
\item\emph{Criminal threats (motive)}. A hacker (source) could attack a car's ECUs (target) by exploiting weakness in the wireless interfaces (vector) to cause a collision or loss of control (consequence)~\cite{Checkoway2011Comprehensive}.
\item\emph{Privacy threats (motive)}. A hacker (source) might be able to intercept private conversations in a car (target) by exploiting vulnerabilities in the TCU (vector) resulting in privacy invasion (consequence)~\cite{Checkoway2011Comprehensive}.
\item\emph{Tracking threats (motive)}. A hacker, or a law enforcement agent, (source) could track a car (target) by exploiting the GPS navigation system (vector), which is mainly used for guiding and directing drivers, resulting in privacy violations~\cite{Brooks2008Automative,Checkoway2011Comprehensive}.
\item\emph{Profiling threats (motive)}. Cars manufacturers (source) can covertly gather cars' logs stored in ECUs (vector) to reveal some driving habits and traffic violations (target) without drivers' consent or knowledge which is a violation of privacy (consequence)~\cite{Brooks2008Automative,Hoppe2011Security}. Another threat is that manufacturers (source) could gather driving habits information (target) that could provide insurance companies with a pool of personal information that might be useful for insurance plans' customization or accident investigations~\cite{Brooks2008Automative}.
\item\emph{Politically-motivated threats (motive)}. A hostile nation (source) might initiate a cyberwar against national transportation roads and their commuters (target) by compromising smart cars that are vulnerable to full remote control (vector) potentially causing large scale collisions and critical injuries (consequence)~\cite{Checkoway2011Comprehensive}.
\end{itemize} 

\section{CPS Security Vulnerabilities}\label{vulns}
In this section, we first highlight the causes of existing vulnerabilities in general CPS. Then we identify application-specific vulnerabilities. For example, not all vulnerabilities found in smart grids are found in medical devices and vice versa. Therefore, we need to distinguish between the generic and application-specific vulnerabilities, so suitable solutions can be designed accordingly.

In addition, using the abstract CPS model proposed in Section \ref{overview}, we classify the vulnerabilities into three types, according to the CPS aspect a vulnerability appears in: \emph{cyber}, \emph{cyber-physical}, and \emph{physical vulnerabilities}. One type of vulnerabilities may appear in different categories. For example, the communication between CPS and the external world (e.g., remote control centers) are considered cyber-, while the communication among CPS components are considered cyber-physical-. Cyber and cyber-physical-vulnerabilities in communication systems usually demonstrate different appearances and properties due to the differences in their origination.

\subsection{Causes of Vulnerabilities}
\vspace{1mm}\noindent\textbf{Isolation assumption}.
The trend of ``security by obscurity" has been dominant in most, if not all, CPS applications since their initial design. The focus has been on designing reliable and safe systems, whereas the security has not been of a great importance. This is because the systems were supposed to be isolated from the outside world, and therefore, considered secure. For example, in ICS and power grids (before they became ``smart"), security relied on the assumption that systems are isolated from the outside world, and the monitoring and control operations were performed locally~\cite{cardenas2008research,Ericsson2010Cyber,Luallen2011Sans}.  Furthermore, medical devices, such as IMDs, were originally designed to be isolated from networks and other external interactions~\cite{Halperin2008Security}. In addition, the same isolation assumption is also present in smart cars where the security of the ECUs' intercommunications relies on their isolation from adversaries~\cite{Larson2008Securing}. Recent and ongoing advances in CPS applications do not adhere to the isolation assumption, but rather more connectivity has been introduced. More connectivity increases the number of access points to cars, thus more attack surfaces arise.

\vspace{1mm}\noindent\textbf{Increased connectivity}.
CPS are more connected than ever before. Manufacturers have improved CPS by adding services that rely on open networks and wireless technologies. For example, ICS and smart grids are connected to control centers which are connected to the Internet or some business-related networks. In fact, most ICS attacks have been internal until 2001; after that most of the attacks originate from external (Internet-based) sources~\cite{Byres2004Myths}. This is clearly due to the increased connectivity deployed in ICS.  In addition,  for fast incident response and more convenience, some field devices are directly connected to the Internet~\cite{Leverett2013Vulnerability,Slay2007Lessons}. Medical devices have wireless capabilities for easier reconfiguration and monitoring. Smart cars have more connectivity so they are referred to as ``connected cars". This connectedness relies on wireless communications such as Bluetooth, cellular, RFID, and satellite radio communications.


\vspace{1mm}\noindent\textbf{Heterogeneity}. CPS consist of components that are usually heterogeneous such that COTS, third party, and proprietary components are integrated to build a CPS application. CPS are almost always multi-vendor systems, and each product has its own security problems. For example, a component might have been manufactured, specified, or implemented by different entities, and eventually integrated by the system deployers. Hence, the building components of CPS are more integrated rather than designed~\cite{Ericsson2010Cyber}. This integration invites the inherent vulnerabilities of each product~\cite{Amin2013IEEENet}. For example, one step of the Stuxnet attack was to exploit the default password in Siemens PLC to access a computer running a Windows OS~\cite{Miller2012SIGITE}. Last, the internal details of the integrated, heterogenous components are unknown, and thus they may produce unexpected behavior when they are deployed. In fact, most of the bugs that led to successful attacks in smart cars, for example, were found at the boundaries of interconnected components manufactured by different vendors, where the incorrect assumptions interact.


\subsection{Cyber Vulnerabilities}

\noindent\textbf{ICS Vulnerabilities}.
\vspace{1mm}

\noindent\textbf{[ICS V1] Communication vulnerabilities in cyber-components}. ICS' reliance on open standards protocols, such as TCP/IP and ICCP, is increasing. This makes ICS applications vulnerable as a result of using vulnerable protocols. TCP/IP's vulnerabilities have been studied and investigated for many years, but the protocol still has security issues as it was not intended to be secure by design~\cite{Bellovin1989Security,Harris1999Tcp}. Another protocol is Remote Procedure Call (RPC), which also has a number of security vulnerabilities, one of which contributed to the well-known Stuxnet attack~\cite{Ms2008Server}. In addition, ICCP, which interconnects control centers, lacks basic security measures such as encryption and authentication~\cite{Nicholson2012ComandSec}.

Wired communications in ICS includes fiber-optic and Ethernet. Ethernet is usually used for local area networks in substations. Because the communication using Ethernet uses the same medium, it is vulnerable to interception and man-in-the-middle (MITM) attacks~\cite{Francia2012CISSE}. For example, an inside attacker could exploit the privilege of accessing the local network and impersonate legitimate components. It is also possible to inject false data or disclose classified information~\cite{Paukatong2005SCADA,Wang2013Cyber}.


Short-range wireless communications are usually performed within the ICS plant assuming an adversary is not able to get close enough to capture wireless traffic. However, the traffic is still vulnerable to be captured, analyzed or manipulated by malicious insiders or even a capable-enough outsider~\cite{DAmico2011Integrating}. In addition, when employees connect their own, probably unsafe, devices to the ICS wireless network, they expose the network to potential threats, so that an attacker could use the employees' personal computers as an attack vector~\cite{Francia2012ACMSE}. Long-range wireless communications such as cellular, microwave, and satellite are also used in ICS. In the literature, long-range wireless communication vulnerabilities have not been studied in the context of ICS. In conclusion, wireless networks are more vulnerable to cyber attacks, including passive and active eavesdropping, replay attack, unauthorized access and others discussed thoroughly in the literature as in~\cite{welch2003wireless,karygiannis2002wireless}.

\noindent\textbf{[ICS V2] Software vulnerabilities}. A popular web-related vulnerability is SQL injection, where attackers can access databases' records without authorization~\cite{Paukatong2005SCADA,Zhu2011ITHINGSCPSCOM}. Databases that are connected directly or indirectly to ICS servers contain important data such as historical data and users' information. Furthermore, emails can also contribute to malware spreading to the network. In~\cite{Fovino2009Experimental}, a number of email-based attacks are shown by experimentation. In addition, gathering security credentials for ICS-connected computers is a very enticing goal for attackers interested in gaining access to a secured network. As a result, the network and the ICS as a whole could be at risk. Finally, vulnerabilities in Internet-exposed devices that are connected to the local network, such as servers in the control center, employees' portable devices like laptops, and smartphones might be exploited to perform malicious activities that affect the desired operations of the control devices~\cite{cardenas2009rethinking}.


\vspace{2mm}\noindent\textbf{Smart grids vulnerabilities}.
\vspace{1mm}


\noindent\textbf{[SG V1] Communication vulnerabilities in cyber-components}. The smart grid's information infrastructure relies on a number of standardized Internet protocols with known vulnerabilities that could be used to launch attacks on the grid. TCP/IP is used for general-purpose connection to the Internet and is not supposed to connect to control centers. However, Internet-faced networks are sometimes connected, directly or indirectly, to the smart grids' control centers due to a network misconfiguration~\cite{Cleveland2012IEC}. This connectivity itself is considered a vulnerability, let alone vulnerabilities in the open protocols. In addition, ICCP, which is the standardized protocol for data exchange between control centers, has a critical buffer overflow vulnerability~\cite{Zhu2011ITHINGSCPSCOM}.

\noindent\textbf{[SG V2] Software vulnerabilities}. Almost the same software vulnerabilities in ICS hold in smart grids with others that are smart grids-specific. For example, widespread smart meters that are remotely upgradeable, inviting a critical vulnerability. An attacker can make use of such a feature to cause blackouts by controlling the meters, either from the control center, or the meters individually. This vulnerability can also be exploited by a software bug~\cite{Anderson2010Controls}.
The grids' components now become more accessible in every household, and hence provide a potential access point for malicious attackers~\cite{Mo2012IEEE}. Some vendors leave backdoors in smart meters. Santamarta~\cite{Santamarta2012Here} was able to discover a backdoor in some smart meters that would result in full control of the meter, including pricing modifications. In addition, some smart meters can be connected to via \textit{Telenet} protocol. This vulnerability can also be exploited to affect other smart meters in the grid to launch coordinated attacks.

\noindent\textbf{[SG V3] Privacy vulnerabilities}. A new type of vulnerabilities has emerged as a result of the two-way communications between smart meters at customers' houses and utility companies. Attackers may be able to intercept the vast amount of traffic generated from smart meters and infer private information about customers~\cite{Cho2014Privacy}. The kind of information attackers could be interested in is, for example, daily habits and residences' presence/absence.

\vspace{2mm}\noindent\textbf{Medical devices vulnerabilities}.
\vspace{1mm}

\noindent\textbf{[MD V1] Security through obscurity}. Because of the lack of \textit{mandatory} security standards for manufacturers of medical devices, some resort to designing proprietary protocols and rely on their secrecy as a security measure~\cite{Lee2012Challenges}. This paradigm, a.k.a ``security through obscurity'' has always failed to thwart attackers.

\noindent\textbf{[MD V2] Communication vulnerabilities in cyber-components}. As most medical devices rely on wireless communications, this implies the devices' vulnerability to a range of wireless-based attacks such as jamming, noise, eavesdropping, replay, and injection attacks. Communications between ICDs and their programmers are vulnerable to eavesdropping due to the lack of encryption. Besides this confidentiality violation, the lack of encryption allows replay attacks~\cite{Halperin2008Pacemakers}. In addition, patients with IMDs or wearable devices could be vulnerable to a number of privacy invasion attacks ranging from discovering the existence of the devices, the devices type, to some physiological measures gathered by the device. In addition, if a device's unique information is inferred, a patient could be vulnerable to tracing attacks~\cite{Halperin2008Security}.

\noindent\textbf{[MD V3] Software vulnerabilities}. The role of software has been growing in medical devices, and so has the likelihood of software vulnerabilities. As a result, recalls of medical devices due to software-related defects has increased~\cite{Hanna2011Take,Fu2013Controlling}. Failure of a device due to a software flaw could result in critical health conditions. Furthermore, Hanna et al. ~\cite{Hanna2011Take} presented the first publicly known software security analysis for medical devices. They found that one type of medical devices, namely Automated External Defibrillator (AED), to have four vulnerabilities: arbitrary code execution due to a buffer overflow vulnerability, weak authentication mechanism, improper credentials' storage, unauthorized firmware update due to improper deployment of the Cyclic Redundancy Check (CRC). In addition, certain assumptions by a device designers could lead to undesired consequences. For example, Li et al.~\cite{Li2011Hijacking} show how a CRC check in the code can lead to dangerous attacks such as replaying outdated measures and sending unauthorized commands.

%


\vspace{2mm}\noindent\textbf{Smart cars vulnerabilities}.
\vspace{1mm}

\noindent\textbf{[SC V1] Communication vulnerabilities in cyber-components}. Cellular interfaces usually serve two purposes: (1) enabling hands-free phone calls and (2) enabling manufactures to perform services remotely such as remote diagnostics, software updates, crash response, and stolen car recovery. TCUs provide cars with such cellular communication channels, among others. However, privacy concerns have emerged from using the cellular interface as a tracking tool where both Global Positioning System (GPS) and the microphone are parts of the TCU. This connection reveals a target's whereabouts, or can become a spying tool via eavesdropping on the in-car conversations by exploiting the microphone~\cite{Checkoway2011Comprehensive,Miller2014aSurvey}.

Bluetooth is one of the most vulnerable attack vectors in smart cars~\cite{Miller2014aSurvey}. When a passenger wants to pair a phone with the car, the only authentication measure is a Personal Identification Number (PIN), prompted by the car's Telematics Control Unit (TCU). This measure is insufficient, and attackers can brute-force the PIN, intercept it, or even inject a false PIN by spoofing the Bluetooth's software. In addition, Bluetooth connections could expose the car to traceability attacks if an attacker successfully extracts the Bluetooth's Media Access Control (MAC) address, which is unique and traceable~\cite{Checkoway2011Comprehensive}.

\noindent\textbf{[SC V2] Software vulnerabilities}. Software is at the heart of every ECU, and smart cars' reliance on it has significantly increased. This, in turn, increases the likelihood of software bugs and security vulnerabilities~\cite{Hoglund2004Exploiting}. For example, if a piece of software is vulnerable to malicious code injection, it would expose the car to various attacks depending on the injected software. Jo et. al.~\cite{jo2016vulnerabilities} show how a TCU running on an Android OS was exploited to unlock doors and trace GPS due to vulnerabilities in the OS. On the other hand, the media player has the ability to directly connect to the CAN bus. This implies that any vulnerability in the player can affect other ECUs because of this connection. Checkoway et al.~\cite{Checkoway2011Comprehensive} identified two vulnerabilities: 1) a malicious specially-crafted CD could affect the media player's ECU and ``reflash" it with malicious data, and 2) the media player is vulnerable to an arbitrary code execution, thanks to its ability to parse different media files.
\subsection{Cyber-Physical Vulnerabilities}

\vspace{2mm}\noindent\textbf{ICS Vulnerabilities}.
\vspace{1mm}

\noindent\textbf{[ICS V3] Communication between ICS components}. ICS relies on protocols that used to be proprietary such as Modbus and DNP3 to monitor and send control commands from a control center to sensors and actuators. ModBus protocol, the \emph{de facto} standard for communication in many ICS, lacks basic security measures, so that it is vulnerable to a plethora of attacks. Its lack of encryption exposes the traffic to eavesdropping attacks~\cite{byres2004use,alcaraz2013critical}. It also lacks integrity checks making data integrity questionable~\cite{byres2004use,Fovino2009Experimental}. In addition, no authentication measures are implemented, suggesting the feasibility of manipulating data traveling to actuators to make them act undesirably, or with data coming from sensors so the controllers can be spoofed by false data~\cite{tsang2010cyberthreats,Zhu2011ITHINGSCPSCOM}. Similarly, DNP3 protocol also does not implement any kind of encryption or authentication mechanisms ~\cite{East2009ATaxonomy,Huitsing2008Attack}. It has, however, a simple integrity measure using CRC. Although CRC is relatively simple, it is better than no integrity check altogether, like in Modbus. East et al.~\cite{East2009ATaxonomy} analyzed the DNP3's vulnerability to at least 23 attacks that exploit the absence of encryption, authentication, and authorization.

Direct access to remote field devices such as RTUs and PLCs used in smart grids is also a vulnerability that might be overlooked by smart grids' operators. Some field devices might be left with default passwords~\cite{Mo2012IEEE}. Furthermore, a large number of PLCs were found to be directly connected to the Internet~\cite{Leverett2013Vulnerability}. In fact, Leverett~\cite{Leverett2011Quantitatively} identified 7,500 field devices that are directly connected to the Internet. Those devices are also used in smart grids, thus the same vulnerability is also applicable.

Sometimes,  in case of failure of the primary communications, it is useful to have a secondary communication channel (e.g., dial-up) to access field devices such as PLCs and RTUs. It provides a direct connection with field devices, which in turn are directly connected with the sensors and actuators~\cite{alcaraz2013critical}. This poses an opportunity for attackers to take control over the field devices without the need to exploit other more advanced communication links, especially in the presence of default logins and simple authentication mechanisms.

\noindent\textbf{[ICS V4] OS vulnerabilities}. The operating systems in ICS devices, such as PLCs and RTUs, are Real-Time Operating System (RTOS), and they do not implement access control measures. Therefore, all users are given the highest privileges, i.e., root access. This is fundamentally insecure, and clearly makes the devices vulnerable to various kinds of attacks~\cite{Zhu2011ITHINGSCPSCOM}.

Applications that are used for controlling and monitoring field devices are running on general-purpose OS. If the OS or running software have vulnerabilities, the hosting computers or laptops posses a potential attack vector on the connected field devices and as a result, their physical components. An example of such exploitable vulnerabilities are two Windows OS vulnerabilities that were exploited in the Stuxnet attack. The first one is a vulnerability in the \textit{Print Spooler Service}, which is vulnerable to remote code execution over remote procedure call (RPC)~\cite{MS2010Spool}. This particular vulnerability allowed Stuxnet to copy itself onto the vulnerable computer~\cite{Chen2011Lessons}. The other exploited vulnerability was in \textit{Windows Server Service} that also was vulnerable to remote code execution through sending a specially crafted RPC request~\cite{Ms2008Server}. Using this vulnerability, Stuxnet connected to other computers~\cite{Chen2011Lessons}. In addition, some attacks are realized by exploiting buffer overflow vulnerabilities in the OS running in the control center~\cite{tsang2010cyberthreats,Zhu2011ITHINGSCPSCOM}. Buffer overflow vulnerabilities are the most commonly reported vulnerabilities to ICS-CERT~\cite{CERT2010Common}.

\noindent\textbf{[ICS V5] Software vulnerabilities}. We consider programs running on general-purpose OS for controlling and monitoring controllers as cyber-physical components. An example of such programs is \textit{WinCC}, which is a Siemens software used for controlling PLCs. In the Stuxnet attack, the first step was to target vulnerable computers running WinCC software~\cite{Chen2011Lessons}. A vulnerable computer is exploited so Stuxnet can copy itself onto the computer, it then installs a rogue driver DLL file that is used by both WinCC software and the PLC. Once the driver DLL installed, a rogue code is sent to the PLC. The critical vulnerability in the controller that allows such an action is the lack of digital signature~\cite{Langner2011Stuxnet}. PLCs and other field devices have limited computational capabilities and cannot perform computationally expensive solutions such as cryptographic measures. Another class of cyber-physical software components cover software running on field devices such as PLCs and other controllers. As we pointed out earlier, the presence of COTS products in CPS is one of the contributing factors for the increased number of vulnerabilities. ~\cite{Leverett2013Vulnerability} revealed an authentication vulnerability in a very common COTS product deployed in 200 PLC models. This vulnerability allows an attacker to bypass the authentication and consequently take control of the PLC. The authors conducted multiple scans and discovered a surprisingly large number of PLCs that directly connect to the Internet. In addition, some vendors leave backdoors in some field devices. This makes it possible for attackers to gain access and full control over the device when valid credentials are gathered~\cite{Santamarta2012Here}.


\vspace{2mm}\noindent\textbf{Smart Grids Vulnerabilities}.
\vspace{1mm}

\noindent\textbf{[SG V3] Grid communication vulnerabilities}.

The power system infrastructure in smart grids relies on almost the same protocols in ICS, such as Modbus and DNP3, thus the same vulnerabilities still hold in smart grids. In addition, IEC 61850 has also been introduced recently as an advancement of these protocols in substations' communications. The lack of security properties in these protocols has a different impact in the context of smart grids. For example, protocols that lack encryption, make the data in transit vulnerable to eavesdropping, which results in a number of attacks such as the inference of customers' usage patterns~\cite{Mo2012IEEE,Mashima2012RAID}, or even injection of false information due to the lack of authentication~\cite{Wang2013Cyber,Rad2011Distributed}. It is also possible to inject the network with bogus packets that aim to flood it, resulting in a DoS attack; or to inject false information, resulting in decisions based on false information~\cite{Sridhar2012IEEE,Xie2010False}.

In addition, smart grids consist of heterogeneous components run by different entities. For example, a generation plant of a grid interacts with a transmission plant, where the transmission plant, in turn, interacts with a distribution plant, and finally the distribution delivers the electricity to end users. Each type of interaction is usually run and administered by different companies, which introduces vulnerabilities in communication and collaboration~\cite{Fleury2008Towards,Hieb2008Security,Mo2012IEEE}.

\noindent\textbf{[SG V4] Vulnerabilities with smart meters}. Smart meters rely on two-way communications, which contribute to a number of new security concerns about an attacker's abilities to exploit such interaction~\cite{Khurana2010SPIEEE}. For example, a smart meter may have a backdoor that an attacker could exploit to have full control over the device. Santamarta~\cite{Santamarta2012Here} analyzed a smart meter's available documentation and found out that there is a ``Factory Login" account. Aside from the customers' accounts with limited capabilities used for basic configurations, this factory login account gives full control to the user over the smart meter. What's more, the communication is transmitted through \textit{telnet} which is known for major security weaknesses, e.g., sending data in clear text without encryption.

Once full control over the smart meter is gained, three potential attacks arise: (1) power disruption, either directly, by malicious interactions with other devices to change their desired power consumption, or indirectly, by injecting false data in a way that the control center receives false information and consequently makes wrong decisions; (2) using the meter as ``bot" to launch attacks possibly against other smart meters or systems within the smart grid network; and (3) the meter's collected data could be tampered with so that the bill reflects false data to reduce the cost to the consumer~\cite{Santamarta2012Here}.

\vspace{2mm}\noindent\textbf{Medical Devices Vulnerabilities}.
\vspace{1mm}

\noindent\textbf{[MD V4] Communication vulnerabilities in cyber-physical-components]}.
The reliance on wireless communications in wearable and IMDs invites vulnerabilities that could result in  physical impacts on patients when exploited. If medical devices fail to transmit or receive expected packets, an undesired health condition could result from incorrect operations performed by the medical device.


As the devices rely on wireless communications, likelihood of jamming attacks arises. For example, when an insulin pump does not receive periodic updates from the associated CGM, it assumes the patient's condition is stable, and no need for an injection of an insulin dosage. This leads to an undesirably high glucose level~\cite{Li2011Hijacking,Radcliffe2011Hacking}. Some devices are vulnerable to battery exhausting attacks, where an adversary exhausts the devices computational or communication resources to withdraw the battery's reserves~\cite{Halperin2008Pacemakers,Rushanansok2014SoK}.

By injecting a specially-crafted packet, it is possible to send unauthorized commands or false data . Halperin et al.~\cite{Halperin2008Pacemakers} and Gollakota et al.~\cite{Gollakota2011They} demonstrated the ICDs' vulnerabilities to injection attacks by exploiting wireless vulnerabilities. In addition, Li et al.~\cite{Li2011Hijacking} demonstrated the insulin pump's vulnerability to be remotely controlled by intercepting the device's communications with its remote control.  Radcliffe~\cite{Radcliffe2011Hacking} also uncovered a vulnerability in the insulin pump device that would allow injection attacks. The device requires its serial number to be part of the command packet as an authentication measure. An attacker equipped with the serial number can inject unauthorized commands to the device.

Replay attacks do not require knowledge of the underlying protocols; instead, an attacker only needs to capture legitimate measurements or command packets, and retransmits them at a later time. Li et al.~\cite{Li2011Hijacking} revealed a vulnerability in an insulin pump that would allow replay attacks so that the pump receives a dishonest reading of the glucose level. And therefore the patient decides mistakenly to inject the wrong amount of insulin such that the decision might threaten the health condition. In addition, Radcliffe~\cite{Radcliffe2011Hacking} revealed that a CGM device was vulnerable to replay attacks. By retransmitting pre-captured packets to the CGM, the author was able to spoof the CGM with incorrect values. In addition, besides the violation of confidentiality, lack of encryption allows replay attacks~\cite{Halperin2008Pacemakers}.


\noindent\textbf{[MD V5] Device authentication}. An attacker can use a commercial programmer without authorization as a result of the implicit trust given to anyone uses the programmers~\cite{Halperin2008Pacemakers}. This makes medical devices vulnerable to safety-critical attacks even without technical knowledge needed for attackers. In addition, some attacks do not need programmers. Instead, Universal Software Radio Peripheral (USRP) is sufficient to replace a programmer and send malicious packets as Halperin et al. ~\cite{Halperin2008Pacemakers} have shown.


\vspace{2mm}\noindent\textbf{Smart Cars Vulnerabilities}.
\vspace{1mm}

\noindent\textbf{[SC V3] Communication vulnerabilities in cyber-physical-components}.
Smart cars are vulnerable to many attacks due to the lack of security considerations in their design~\cite{Larson2008Securing}. In-vehicle communication protocols, such as CAN and LIN, lack encryption, authentication and authorization mechanisms. Here we review the vulnerabilities in the most common protocol, CAN. The CAN protocol has a number of vulnerabilities that contribute to most of the attacks on smart cars. For example, CAN protocol lacks critical security properties such as encryption, authentication, and has weak authorization and network separation. Due to the lack of encryption, TPMS is vulnerable to eavesdropping and spoofing ~\cite{ishtiaq2010security}. Tracing a car is possible by exploiting the unique ID in the TPMS communications. In addition, the CAN protocol's broadcast nature increases the likelihood of DoS attacks~\cite{Experimental2010Koscher}. CAN bus error handling mechanism makes it vulnerable to DoS~\cite{cho2016error}. Another security property, common in computer security literature, is non-repudiation, where there is no way to identify the origin of a particular message~\cite{Hoppe2011Security}. Clearly, these vulnerabilities, especially the lack of encryption and authentication, result from the isolation assumption discussed earlier.

\noindent\textbf{[SC V4] Comfort ECUs}.
More advanced features are continuously added to ECUs to improve safety and comfort. For example, ECUs like \textit{Adaptive Cruise Control} (ACC), \textit{Lane Keep Assist}, \textit{Collision Prevention} provide safety, where \textit{Comfort Park Assist} and RKE are examples of ECUs that provide comfort. Although these components have a great impact on improving the driving experience in terms of safety and comfort, they pose a new type of attacks, i.e., cyber-physical attacks. These components are part of the CAN network, and there is a threat of attacking them and compromising their expected functions by directly exploiting their vulnerabilities, or vulnerabilities in other ECUs residing in the same network~\cite{Miller2014aSurvey}. As an example, ACC is the next generation of \textit{cruise control}. It provides the ability to detect the speed of the car ahead  using laser or radar sensors, and adaptively change the current speed to maintain a safe distance between cars. A well-equipped attacker might be able to interrupt ACC sensors' operations by either introducing noise or spoofing. As a result of the attack, ACC may reduce or increase speed unexpectedly, or even cause collisions. The threat is the ability to tamper with the sensors externally, or with the ACC's ECU itself internally, possibly through other ECUs that are potentially vulnerable to remote attacks such as TPMS, RKE, or TCU.

%

\noindent\textbf{[SC V5] Vulnerabilities with X-by-wire}.
An emerging trend in smart cars is the ``X-by-wire". It aims to gradually replace the mechanically-controlled components in the car, such as the steering wheel and the brake pedal, by electronic or electro-mechanical components. Such electro-mechanical components would make drivers control the relevant functionality by only pressing some buttons. Steer-, Drive-, Brake-, Shift-, and Throttle-by-wire are all examples of this trend~\cite{Studnia2013Survey}. This implies new opportunities for attackers to launch cyber-physical attacks exploiting such new functionalities. However, this technology relies on FlexRay communication protocol, which is more advanced than CAN protocol in terms of speed and safety features. However, it is more costly and less likely to widespread in the near future~\cite{Studnia2013Survey}.
\subsection{Physical Vulnerabilities}

Finally, we review vulnerabilities in physical components that would cause cyber impact. Physically tampering with a physical component or its surrounding environment could result in misleading data in the cyber-physical components. Most of the vulnerability analysis in CPS literature focus on cyber attacks with physical impact. Very few, on the other hand, studies physical attacks with cyber impact such as in~\cite{MacDonald2013Cyber}.

\vspace{2mm}\noindent\textbf{ICS Vulnerabilities}. The physical exposure of many ICS components, such as RTUs and PLCs, that are scattered over a large area is a vulnerability in itself. With insufficient physical security provided to these components, they become vulnerable to physical tampering or even sabotage. For example, a water canal's sensors rely on solar panels as a source of energy so they can communicate with the control center. These panels were stolen, and therefore, the control center lost critical data necessary for the desired operations~\cite{Amin2010Stealthy}.

\vspace{2mm}\noindent\textbf{Smart Grids Vulnerabilities}. Similar to ICS, smart grids' field devices are placed in unprotected environments. A huge amount of physical components are highly exposed without physical security, and thus vulnerable to direct physical destruction. For example, power lines are vulnerable to malicious, accidental, and natural attacks. For example, in Northern Ohio, overgrown trees caused a large blackout affecting over 50 million people~\cite{tsang2010cyberthreats}.

 In addition, smart meters attached to buildings, houses, and remote areas make them an easy target to various physical attacks. Mo et al.~\cite{Mo2012IEEE} suggest that it is even infeasible to achieve adequate physical protection of all assets in smart grids. Therefore, it is necessary to devise prevention and detection solutions.

\vspace{2mm}\noindent\textbf{Medical Devices Vulnerabilities}. Medical devices, whether implantable or wearable, sometimes could be vulnerable to physical access. That is, the attacker's ability to physically deal with the medical device. For example, for maintenance purposes, an attacker exploiting the absence of the device's owner, tampers with it, and potentially performs malicious activities such as malware installation, or modification of the configurations such that the device delivers unadvised treatment that could threaten the patient's health. In addition, by physical access to a device, it is also possible to get the device's serial number, which is useful for some attacks~\cite{Radcliffe2011Hacking}.

In general, physical access to the device opens up many possibilities to various attacks. Hanna et al.~\cite{Hanna2011Take} recommend protecting medical devices from physical access by any potential attacker. Another subtle vulnerability to consider is the mobility of medical devices' users. It could be a vulnerability by itself. As the device's designers cannot control patients' surroundings, the devices could be vulnerable to unpredicted physical attacks when a patient is in an insecure location. This is especially true for politically-motivated attacks~\cite{Lisa2013Doctors}.

\vspace{2mm}\noindent\textbf{Smart Cars Vulnerabilities}. If not physically protected, cars can be vulnerable to numerous attacks that do not necessarily require cyber-capabilities. For example, TPMS parts could be destroyed resulting in DoS attack such that TPMS cannot send tires' air pressure to the designated ECU. In addition, exposing the car to any kind of physical access is another vulnerability that could cause critical attacks. For example, a mechanic can get physical access to a car's internal parts directly through OBD-II port~\cite{Wolf2004Securityin}. Furthermore, some external parts can be used to access critical components in the car's CAN network. Some attackers could get access to the internal network through exterior exposed components such as the exterior mirrors~\cite{Hoppe2011Security}.

\begin{table}[ht!]
	\begin{footnotesize}
		\begin{tabular}{|p{1.3cm}p{4cm}cc|}
			\hline
			\textbf{CPS} & \textbf{Vulnerability}& \textbf{Type} & \textbf{Cause}\\\hline
			ICS &  &  &  \\
			& Open communication protocols & C & I, O \\
			& Wired communications & C & I, H, S \\
			& Wireless communications & C & I, Con \\
			& Web-based attacks &  C & Con, H  \\
			& Insecure protocols & CP & I, Con \\
			& Interconnected \& exposed field devices & CP & I, Con \\
			& Insecure secondary access points & CP & I, Con \\
			& Insecure OS \& RTOS & CP & I, Con, H\\
			& Software & CP & Con, H \\
			& Equipments' physical sabotage & P & I \\
			\hline
			Smart Grid &  & &  \\
			& Blackouts & CP & I, Con, H \\
			& Communication protocols & C & I, Con \\
			& Software & C & I, Con, O, S\\
			& Customers' privacy invasion & C & I, Con, H \\
			& Interconnected field devices & CP & I, Con, H\\
			& Insecure protocols & CP & I, Con \\
			& Insecure smart meters & CP & I, Con, H \\		 	 			
			& Equipments' physical sabotage & P & I \\
			\hline
			Wearable \& IMDs &  &  & \\
			& Jamming \& noise& P  & Con\\
			& Replay \& injection attacks & C, CP  & I, Con\\
			& Patient's privacy invasion & C  & I, Con\\
			& Software & C, CP  & I, H\\
			& DoS & CP & Con\\
			\hline
			Smart cars & & & \\
			& TPMS easy interception & CP & H \\
			& GPS traceability & C & H \\
			& Bluetooth authentication flaw& C & Con \\
			& Insecure CAN bus &  CP & I,  Con  \\
			& Replay attacks & CP & I, Con \\
			& Communication software flaws & C & Con \\
			& Media player exploitations & C & H \\
			& Physically unprotected components & P    &  I \\
			\hline
		\end{tabular}
		\caption{Summary of vulnerabilities. C: Cyber, CP: Cyber-Physical, P: Physical; I: Isolation assumption, C: Connectivity, O: Openness, H: Heterogeneity, S: Many stakeholders.}\label{table2}
	\end{footnotesize}\vspace{-5mm}
\end{table}

\section{Real-world CPS attacks}
\label{attacks}

We review reported cyber, cyber-physical, and physical attacks on the four CPS applications that exploited the aforementioned vulnerabilities in Section~\ref{vulns}. In general, publicly known attacks are rare~\cite{Pietre2011Cybersecurity}, and it is infeasible to find attacks that represent exploitations of all vulnerabilities in section ~\ref{vulns}. Instead, we consider attacks that have been realized by experimentation or in real life. At the end of this section, we describe the discussed cyber-physical attacks using a cyber-physical taxonomy proposed by Yampolskiy et al.~\cite{Yampolskiy2013HCNS} in tables \ref{ICSattacks}, \ref{SmartGridAttacks}, \ref{MedicalAttacks}, and \ref{CarsAttacks} for attacks on ICS, smart grids, wearable and IMDs, and smart cars, respectively. Their proposal dissects CPS attacks into six-dimensional description, by which allows us to gather more insights about each attack. The description includes the attacked object (Influenced Element), the resulting changes on the attacked object from the attack (Influence), indirectly affected components (Affected Element), changes on the CPS application (Impact), how the attack took place (Method), and preceding attacks needed to make an attack successful (Precondition). We also integrate our CPS framework into this taxonomy by highlighting CPS aspects in the attack tables with \textit{C}, \textit{CP}, and \textit{P}, for cyber, cyber-physical, or physical aspects, respectively.

In this section, we categorize the attacks based on the damages' location. Attacks that do not reach sensors/actuators are considered purely cyber, while attacks that directly impact physical components are physical. Whereas, attacks that indirectly impact physical components, through cyber components, are cyber-physical.

\subsection{Cyber Attacks}
\vspace{2mm}\noindent\textbf{ICS Attacks}

\emph{Communication protocols}.
A number of attacks have exploited vulnerabilities in communication protocols. For example, spoofing attacks on Address Resolution Protocol (ARP) were demonstrated on SCADA system~\cite{Urias2012MILCOM,Francia2012ACMSE}.

\emph{Espionage}.
\textit{DuQu} and \textit{Flame} are two examples of ICS attacks with spying purposes~\cite{Duqu2012Chien,Munro2012Deconstructing}. Flame, for example, targeted various ICS networks in the Middle East and was discovered in 2012. This malware's main goal was to collect corporations' private data such as emails, keyboard strokes, and network traffic~\cite{Munro2012Deconstructing}. Although the intention of the attack was not clear, hostile nations or industrial competitors could benefit from such information leakages.

In addition, in 2013, a group of hackers, known as \textit{Dragonfly}, targeted energy firms in the U.S. and Europe. The attackers main goal seems to have been gathering private information. To do that, they needed to infect systems in the targeted firms with malware that grants remote access. They started by sending phishing emails to the personnel of the targeted firms containing malicious PDF attachments. Then the attack vector escalated to exploiting \textit{watering hole} vulnerabilities in victims' browsers by directing victims to visit malicious websites hosted by the attackers. Both delivery mechanisms infected targeted systems with a malware that allowed attackers to gather private information in the infected systems~\cite{Symantec2014Dragonfly}.

\emph{Unintentional attack}. Although software updates are critical to ensuring systems are vulnerability-free, it can be a source of service disruption. For example, in a nuclear plant, one computer in the control center was updated and rebooted thereafter. The reboot erased critical data on the control system. As the data was erased, other components of the system misinterpreted such loss, resulting in abnormality in operations and the plant's shutdown~\cite{cardenas2008research}.

\emph{Web-based attacks}. \textit{Night Dragon} attack, in 2011, targeted sensitive information from private networks of a number of energy and oil companies~\cite{alcaraz2013critical}. The attack combined a number of web-related vectors to succeed such as SQL injection vulnerability and a malware injection~\cite{alcaraz2013critical,Miller2012SIGITE}.



\vspace{2mm}\noindent\textbf{Smart Grids Attacks}


\emph{DoS}.
 The traffic in smart grids is time-critical, so delays may result in undesired consequences. Flooding the network at different layers is the probable approach to achieve DoS attacks. Lu et al.~\cite{Lu2010Review} evaluated the impact of DoS on smart grids' substations. The authors found that the network performance only gets affected if the flooding is overwhelming. In addition, at the physical layer, the deployment of wireless communications increases and therefore, jamming attacks' are possible as shown in~\cite{Lu2011Jammer}.

\emph{False data injection}.
Introducing false data in smart grids' traffic leads to different consequences such as service disruption and financial losses. Liu et al.~\cite{Liu2011False} demonstrated a simulated false data injection to evaluate the impact on the state estimation in smart grids. The authors assumed the attacker's pre-intrusion to the control center for a successful attack, which aimed to ultimately inject false measurements to smart meters to disrupt the state estimate process. Such disruption leads to financial losses for the operating utilities~\cite{Wang2013Cyber}.

\emph{Customers' information}.
Attackers can analyze network traffic in smart grids between smart meters and data centers to infer private information about customers. For example, an attacker can determine if a target is available at home at particular times and dates. In addition, it is also possible to deduce lifestyle in terms of sleeping times and quality, preferred home appliances, and many more~\cite{Molina2010Private}.

\emph{Untargeted malware}.
In 2003, the Slammer worm resulted in disabling the traffic between field devices and substations. Although that malware was not intended to affect the energy sector, it still had an effect because of the interconnectedness of the smart grid networks. The malware consumed a significant amount of the time-critical traffic, but did not cause service outages~\cite{Byres2004Myths}.


\vspace{2mm}\noindent\textbf{Medical Devices Attacks}

Most, if not all, of the reported attacks on medical devices have been performed in experimental environments. However, the possibility of the attacks in this section, should raise an alarm to stimulate the efforts in improving security in medical devices.
Although some attacks are specific to certain devices, such as insulin pumps, the same attack techniques could be applicable to other IMDs and wearable devices. This is the case because of the similarities in the communication links and hardware components.

\emph{Replay attacks}.
By exploiting a vulnerability in an insulin pump, replaying eavesdropped packets is possible by incorporating a previously intercepted device's PIN~\cite{Li2011Hijacking}. In addition, replay attacks could result in misinformed decisions regarding insulin injection~\cite{Radcliffe2011Hacking}.

For example, by replaying an old CGM packet to an insulin pump, a patient will receive a dishonest reading of the glucose level, and therefore will decide, mistakenly, to inject the wrong amount of insulin. Such wrong decision could result in critical health condition.

\emph{Privacy invasion}.
Attacks violating patients' privacy have different goals and consequences. For example, for the remote control attack on the insulin pump in~\cite{Li2011Hijacking}, an attacker needs to learn the device type, PIN, and legitimate commands sent from the remote control. The authors successfully performed this attack and revealed three types of privacy-related information, namely, the devices' existence, its type, and finally the PIN. In addition, Halperin~\cite{Halperin2008Pacemakers} demonstrated similar attacks on an ICD medical device such as revealing patient's personal and medical information and the device's unique information.

\vspace{2mm}\noindent\textbf{Smart Cars Attacks}

 We reviewed about two dozen papers that discuss the security of smart cars. Most of the work done is abstract, theoretical, or simulation-based. Only a few present results of actual experiments on real cars~\cite{Hoppe2011Security,Experimental2010Koscher,Checkoway2011Comprehensive,Miller2013Adventures}.

In order for a successful attack on a car, an attacker needs to gain access to the internal network physically, through the OBD-II port, media player, or USB ports, or wirelessly, through the Bluetooth or cellular interfaces. Once an attacker gets into the car's internal network, a plethora of attacks opportunities are open.

\emph{DoS}.
DoS attacks can take on different forms whose impacts vary in safety-criticality. Koscher et al.~\cite{Experimental2010Koscher} disabled CAN communication from and to the Body Control Module (BCM) which resulted in a sudden drop from 40 to 0 MPH on the speedometer.
In addition, this attack also resulted in freezing the whole Instrument Panel Cluster (IPC). For example, if the speedometer was at 60 MPH before the attack, and the driver increased the speed, there will be no change in the speedometer and the driver might reach a dangerous speed level.

\emph{False Data Injection (FDI)}.
An example of this attack is displaying a false speed on the speedometer. An attacker would first intercept the actual speed update packet sent by the BCM, and then transmit a modified packet that had the false speed~\cite{Experimental2010Koscher}. In addition, an attacker can forge the real status of the airbag system to appear healthy, even if the airbag had malfunctioned or was removed~\cite{Hoppe2011Security}. Another type FDI is shown in~\cite{Guan2016From} where the authors showed how a passenger can manipulate the data collected by insurance dongles used for rate customization in a way that would resulted in an undeserved lower insurance price.

\emph{Privacy invasion}.
Checkoway et al.~\cite{Checkoway2011Comprehensive} were able to exploit the cellular interface in the TCU and eavesdrop on in-car conversations. In addition, a report published by Ed Markey, a U.S. senator, reveals that car manufacturers store a large amount of private information such as driving history and cars' performance~\cite{Markey2015Tracking}.

\subsection{Cyber-Physical Attacks}

\vspace{2mm}\noindent\textbf{ICS Attacks}

\emph{Legacy communication channels}.
As we mentioned above, dial-up connections provide direct access to field devices and sometimes their security are overlooked. In 2005, billing information of a water utility was accessed by exploiting the dial-up connection in a canal system ~\cite{turk2005cyber}. Although this attack did not have a physical impact, it could have, due to the control capabilities provided by the dial-up connection.

\emph{Disgruntled insiders}.
A huge financial loss for utility companies with undesired environmental impacts could result from attacking a water and sewage system. In 2000, an ex-employee intentionally disrupted the operations of a sewage treatment system in Maroochy Water Services in Queensland, Australia. The attacker exploited his knowledge, as a previously-legitimate insider, to change configurations in pumping stations using a laptop and a radio transmitter. The consequence of the attack caused a huge amount of raw sewage to flood into the streets and taint the environment~\cite{Slay2007Lessons}.

\emph{Modbus worm}.
Fovino et al.~\cite{Fovino2009Experimental} presented an alarming work on targeted malware. The authors crafted malware that exploits the lack of authentication and integrity vulnerabilities in Modbus protocol. The worm aims to perform two attacks: DoS, by identifying sensors or actuators and sending them DoS-inducing messages, and command injection, by sending unauthorized commands to the sensors or actuators.


\emph{Malware}.
Some malware targets specific systems to achieve goals like traffic interception and interruption of operations. They exploit software vulnerabilities in applications that manage control field devices. A well-known example is \textit{Stuxnet}. This attack is considered one of the most sophisticated attacks on ICS that clearly embodies cyberwar. Stuxnet exploits software vulnerabilities to achieve physical consequences~\cite{Yampolskiy2013HCNS}. Because the targeted networks were off the Internet, it is believed that the delivery mechanism was a USB stick. The attack can be generally summarized  into two phases: 1) spreading and determining targets and 2) PLCs hijacking~\cite{Langner2011Stuxnet}. The first step was realized by exploiting two zero-day Windows vulnerabilities, i.e, one in the shared printing server and the other was in Windows Server Service. Both vulnerabilities would allow remote code execution using RPC. Stuxnet used the first vulnerability to install itself in the system and the other to connect to other systems to also install itself in an iterative fashion. This process led to infecting about 100,000 infected systems worldwide, however, because the attack was targeted on specific PLCs, the infection did not have an influence on systems that were not connected to the targeted PLCs. Once Stuxnet is installed, it looks for a specific software used for monitoring and sending commands to the PLCs, that is Seimens WinCC. It goes through a thorough analysis to ensure that WinCC is connected to one of specific Siemens PLCs~\cite{Langner2011Stuxnet}. Once determined, the malware injects the malicious code that aims to alter PLCs' configuration. Once that is achieved, the final objective of the attack is realized, which is, most likely, damaging centrifuges used for uranium enrichment. For a detailed Stuxnet analysis, we refer you to~\cite{W322011Falliere}.

\emph{Web-based attacks}.
A group of hackers exploit a web-based interface that is directly connected to field devices like PLCs. They opened multiple connections and left them open until authorized users could not access them, resulting in a DoS attack. In addition, they also sent a web page to the controller/ field device that contains malicious Java script code designed to exploit a bug in the TCP/IP stack causing resetting of the controller~\cite{turk2005cyber}.

\vspace{2mm}\noindent\textbf{Smart Grids Attacks}

\emph{Cyber extortion}.
This type of attack is rare, at least publicly, where attackers take control over the target smart grid and make demands as a price of not causing a large-scale blackout~\cite{Nakashima2008Hackers}.

\emph{Blackouts}.
In the context of smart grids, a blackout is considered a DoS attack. The availability of smart grids is probably the most important security goal to maintain, and attacks aiming to compromise it could result in a large-scale blackout that might have a nationwide impact. In 2007, Idaho National Laboratory (INL) demonstrated an experiment on how a generator could be damaged as a result of a cyber attack~\cite{Cnn2007Sources}. The experiment proved the feasibility of such attacks. For example, it is believed that two blackouts in Ohio and Florida in 2003 were caused by a Chinese politically-motived group, the People's Liberation Army~\cite{Harris2008China}. In addition, about 800 blackouts in the U.S. occurred in 2014 for unknown reasons~\cite{Eaton2014Power}. Some speculations suggest that such mysterious outages may have resulted from cyber-physical attacks launched by hostile nations~\cite{Harris2008China}.

\vspace{2mm}\noindent\textbf{Medical Devices Attacks}

\emph{DoS attacks}. This attack, when successful, could lead to critical health condition to patients. Halperin et al.~\cite{Halperin2008Pacemakers} were able to disable an ICD therapies by replaying a previously recorded ``turn off" command sent by the programmer.

\emph{False data and unauthorized commands injection}.
Li et al.~\cite{Li2011Hijacking} were able to remotely control an insulin pump's remote control and successfully stopped and resumed the insulin injection from a 20 meter distance.

\emph{Replay attacks}.
By exploiting a software vulnerability in the replay attacks countermeasure, any packet can be retransmitted to the CGM and insulin pump~\cite{Li2011Hijacking,Radcliffe2011Hacking}.

\vspace{2mm}\noindent\textbf{Smart Cars Attacks}

\emph{DoS}. One form of DoS attacks is where an attacker prevents passengers from closing any opened windows. Another is to disable the warning lights or the theft alarm system, so that the car cannot produce warnings and alarms when needed~\cite{Hoppe2011Security}. In addition, jamming of wireless communications is also a form of DoS such as jamming RKE signals~\cite{Wetzels2014Broken}.

\emph{Malware injection via Bluetooth}.
Checkoway et al.~\cite{Checkoway2011Comprehensive} conducted an attack that exploits compromised devices connected to the car through a Bluetooth connection. The authors assume the attacker's ability to first compromise a connected device to the car, usually a smartphone. Then they launch an attack exploiting the connectivity between the Bluetooth's ECU, which is the TCU, with the other ECUs. This was realized by installing a hidden malware, Trojan Horse, on the connected smartphone. The malware captures Bluetooth connections and then sends a malicious payload to the TCU. Then once the TCU is compromised, the attacker can communicate with safety-critical ECUs, such as the Anti-Lock Braking System (ABS). In addition, Woo et al. ~\cite{woo2015practical} showed a wireless attack that exploits a malicious diagnostic mobile app connected to the OBD-II port via Bluetooth. Since the app runs on a mobile device, te attack can be launched through a cellular network.

\emph{Malware injection via cellular network}.
The cellular channel in the TCU is exploitable and vulnerable to malware injection attacks. The attack is realized by calling the target car and injecting the payload by playing an MP3 file~\cite{Checkoway2011Comprehensive}.

\emph{Malware injection via the OBD-II port}.
Malware injection through the OBD-II port needs physical access to the car. Hoppe et al.~\cite{Hoppe2011Security} show how an injected malware can launch a number of DoS attach such as preventing passengers from opening and closing windows and also preventing the car from displaying that airbags are lost.

\emph{Packet injection}.
This attack requires previous access to the CAN network. Once an attacker gets into the network, physically or wirelessly, a large number of attacks are possible. For example, through the OBD-II port, it is possible to increase the engine's Revolutions Per Minute (RPM), disturb the engine's timing, disable the engine's cylinders, and disable the engine itself. In addition, attacks on brakes are also possible by injecting random packets to the Electronic Brake Control Module (EBCM) such that it locks and releases the brakes resulting in unsafe driving experiences~\cite{Experimental2010Koscher}. In addition, Lee et. al.~\cite{lee2015fuzzing} performed fuzzing attacks on multiple cars that resulted in arbitrary behaviors that could affect passengers' safety. The fuzzing attack is simple, they captured the CAN IDs of the normal network traffic and then flood the network with packets that have the same IDs but different data fields.

\emph{Replay attacks}.
This attack requires two steps: 1) intercepting the CAN network traffic when certain functions are activated and 2) retransmitting the observed packet to reactivate the same function.
Koscher et al.~\cite{Experimental2010Koscher} were successfully able to disable the car's interior and exterior lights by sending previously-eavesdropped packets.


\subsection{Physical Attacks}

\vspace{2mm}\noindent\textbf{ICS Attacks}

\emph{Untargeted attacks}.
Zotob worm, although not targeted on ICS, caused manufacturers to shut down their plants. For example, US-based DaimlerChrysler had to shutdown thirteen of their manufacturing plants for about an hour~\cite{tsang2010cyberthreats}. Such an incident stimulated researchers to examine the impact of unintended malware intended for tradition IT systems, on ICS networks. For example, Fovino et al.~\cite{Fovino2009Experimental} showed how unintended malware could result in collateral damages ranging from causing ICS servers to reboot, opening potential arbitrary code execution vulnerabilities, infecting personal computers, and stimulating DoS attacks.

\vspace{2mm}\noindent\textbf{Smart Grids Attacks}

\emph{Natural and environmental incidents}.
We give a few examples of power blackouts in 2014 that resulted from natural causes to show the impact of physical exposure and unreliability of smart grids' components. An ice storm in Philadelphia affected 750,000 people for several days with no electricity, whereas a tornado hit the New York area affecting 500,000 people~\cite{Eaton2014Power}. Furthermore, the widespread power transmission lines around various environments and conditions contributed to unexpected attacks such as overgrown or falling trees. For example, some overgrown trees caused a large blackout affecting over 50 million people in Northern Ohio~\cite{tsang2010cyberthreats}. This incident, however, is controversial and security analysts suggest that it resulted from a cyber-physical attacks originating from China~\cite{Harris2008China}. In addition, in 2014, wild animals caused 150 blackout in the U.S. by eating and damaging cables~\cite{Eaton2014Power}.

\emph{Theft}.
Copper wires and metal equipments are profitable targets for financially-motivated thieves. For examples, theft caused a blackout with an impact on 3,000 people in West Virginia~\cite{Eaton2014Power}.

\emph{Car accidents}.
In 2014, 356 outages in the U.S. were caused by cars hitting transmission towers, transformers, or power poles~\cite{Eaton2014Power}.

\emph{Vandalism}.
Attackers can physically damage parts of smart grids such as cables, poles, generators, smart meters, and transformers. An example of that is an incident in 2013 where a sniper in California shot more than a hundred shots at a transmission substation, leaving 17 transformed damaged~\cite{Fox2014Threat}.

\emph{Terrorist attacks}.
In 2014, the first terrorist attack on a power grid occurred in Yemen. The attackers launched rockets to destroy transmission towers and caused a nationwide blackout affecting 24 million people~\cite{InvestmentWatch2014First}.

\vspace{2mm}\noindent\textbf{Medical Devices Attacks}

\emph{Acquiring unique IDs}. Obtaining devices' serial numbers is an example of attacks that require physical access to the target devices~\cite{Radcliffe2011Hacking}.

\vspace{2mm}\noindent\textbf{Smart Cars Attacks}

\emph{Relay attacks}.
This kind of attack targets the RKE, where an attacker relays the communications between the car and its key fob. The attack exploits the periodical LF beacon signal the car sends to detect if the key fob is in close range. The attacker captures and relays it, using an antenna, to the relatively far key fob, which is most likely in the car owner's pocket. The key fob is activated by the relayed signal and sends an ``open" Ultra High Frequency (UHF) signal to open the car. Once the car is opened, the same attack is repeated from inside the car to start it. The attack was implemented successfully on ten different cars from eight manufacturers. In addition, it  evades cryptographic measures because it targets communications at the physical layer~\cite{Francillon2011Relay}. This kind of attack is also called Man-in-the-Middle (MITM) or ``two-thief" attacks~\cite{Wetzels2014Broken}. In addition, Garcia et al.~\cite{garcia2016lock} presented a attack on the RKE system used in many cars. The attack relies on exploiting the simple cryptographic algorithms and inadequate key management in order to clone a car's key resulting in an unauthorized access to the car.
key

\emph{ABS Spoofing}.
Shoukry et al.~\cite{Shoukry2013Non} showed a physical attack on the ABS wheel speed sensor. The sensor measures the speed of a wheel using a magnetic field the gets induced by the iron tone wheel attached to every wheel. In the demonstrated attack, the authors introduced a malicious actuator that produces another magnetic field that disrupts the original field produced by the wheel speed sensor. This additional field results in inaccurate speed measurements received by the ABS ECU. The authors also demonstrated a spoofing attack that allows attackers to inject incorrect speed measures and spoof the ABS ECU.


\begin{table*}[t]
	\begin{footnotesize}
	\begin{tabular}{|p{1.3cm}p{1.5cm}p{2cm}p{2cm}p{2cm}p{2cm}p{2.5cm}p{1.5cm}|}
		\hline
		\textbf{Name} & \textbf{IE} \footnote{Influenced Element} & \textbf{I} \footnote{Influence} & \textbf{AE} \footnote{Affected Element} & \textbf{Impact} & \textbf{Method} & \textbf{Precondition} & \textbf{Ref} \\
		\hline
		Maroochy & Pumps & Pumps work \newline undesirably & Correct settings in pumping stations manipulated &  Raw sewage flood in streets, tainted environment, and financial losses & Used a laptop and a radio transmitter to manipulate pumps & Insider knowledge & ~\cite{Slay2007Lessons} \\
		\hline
		Modbus worm & ICS network & Infected ICS \newline network & Connected field devices & Servers' rebooting, DoS, \& unauthorized commands injection & Inject malware code into ICS network traffic & Access to ICS traffic & ~\cite{Fovino2009Experimental}                     \\
		\hline
		Stuxnet & Centrifuges PLCs    & Exaggerated rotation of centrifuges & Centrifuges' \newline rotors & Lifetime reduction \& physical damage  & Illegitimate commands from PLCs sent to centrifuges & Infected PLC by Stuxnet & ~\cite{Yampolskiy2013HCNS,Chen2011Lessons,Langner2011Stuxnet,W322011Falliere} \\
		\hline
		Web-based attacks & Field devices (e.g. PLCs) & Field devices web interface feature & The physical \newline environments controlled by devices & Legitimate personnel unable to connect to field devices remotely or locally (DoS) & Leave devices with open connection state & Devices are directly exposed to the Internet & ~\cite{turk2005cyber}                     \\
		\hline
		Web-based attacks & Field devices (e.g. PLCs) & Field devices web interface feature & The physical \newline environments controlled by devices & Devices lost configurations & Malware injection & TCP/IP vulnerability in a COTS implementation & ~\cite{turk2005cyber}                     \\
		\hline
	\end{tabular}
\caption{ICS Cyber-Physical attacks}
\label{ICSattacks}
\end{footnotesize}
\end{table*}

\begin{table*}[t]
	\begin{footnotesize}
	\begin{tabular}{|p{1.3cm}p{1.5cm}p{2cm}p{2cm}p{2cm}p{2cm}p{2.5cm}p{1.5cm}|}
		\hline
		\textbf{Name} & \textbf{IE}  & \textbf{I} & \textbf{AE} & \textbf{Impact} & \textbf{Method} & \textbf{Precondition} & \textbf{Ref} \\
	     \hline
		Cyber extortion & Power delivery & Utilities lose control over their grid system & Customers  & Lost of services \& financial losses & Exploit Internet-connected smart grid components & Inside knowledge & ~\cite{Nakashima2008Hackers} \\
		\hline
		Aurora experiment & Circuit breakers (P) & Change relay behavior (CP) & Power generators and power-fed substations (CP) & Physical damage to generators \& inability to deliver electricity (P) & Unexpected opening \& closing of circuit breakers (CP) & Access \& inside knowledge (CP) & ~\cite{Zeller2011Myth}                     \\
		\hline
	\end{tabular}
	\caption{Smart Grids Cyber-Physical attacks}
	\label{SmartGridAttacks}
	\end{footnotesize}
\end{table*}

\begin{table*}[t]
	\begin{footnotesize}
	\begin{tabular}{|p{1.75cm}p{1.5cm}p{2cm}p{2cm}p{2cm}p{2cm}p{2.5cm}p{1cm}|}
		\hline
		\textbf{Name} & \textbf{IE}  & \textbf{I} & \textbf{AE} & \textbf{Impact} & \textbf{Method} & \textbf{Precondition} & \textbf{Ref} \\
		\hline
		DoS &  A medical device    &  The device is turned off  &      Patients   & Patient does not receive expected therapy (DoS)  &    Retransmit ``turn off" command   &  Capture ``turn off" previously sent by the programmer   &     ~\cite{Halperin2008Pacemakers}   \\
		\hline
		False data injection (FDI) & Insulin pump &  False measurements sent to insulin pump &  Patient's therapeutic decisions &  Wrong decisions &  Impersonate CGM and by sending similar packets with false data &  Interception of CGM and insulin pump communications  &  ~\cite{Li2011Hijacking}   \\
		\hline
		Unauthorized commands injection & Insulin pump  &  Unauthorized commands sent to insulin pump &  Patient's safety &  Dangerous health condition &  Impersonate insulin pump remote control by sending similar packets with unauthorized commands &  Interception  communications between insulin pump and its \newline remote  &  ~\cite{Li2011Hijacking}   \\
		\hline
	\end{tabular}
	\caption{Medical devices Cyber-Physical attacks}
	\label{MedicalAttacks}
	\end{footnotesize}
\end{table*}

\begin{table*}[t]
		\begin{footnotesize}
	\begin{tabular}{|p{1.5cm}p{1.5cm}p{2cm}p{2cm}p{2cm}p{2cm}p{2.5cm}p{1.5cm}|}
		\hline
		\textbf{Name} & \textbf{IE}  & \textbf{I} & \textbf{AE} & \textbf{Impact} & \textbf{Method} & \textbf{Precondition} & \textbf{Ref} \\
		\hline
		DoS 1& BCM & Sudden drop in speedometer & IPC & Frozen IPC and failure in turning car on/off & Disabling communication to/from BCM & Physical access to CAN bus & ~\cite{Experimental2010Koscher}        \\
		\hline
		DoS 2& Windows &  Window closing  failure & Windows control buttons & Discomfort and frustration & Reverse engineering and fuzzing & Physical access & ~\cite{Experimental2010Koscher,Hoppe2011Security}             \\
		\hline
		Malware \newline injection 1	&    Bluetooth ECU       &     Ability to connect to other ECUs     &   Safety-critical ECU with cyber-phyiscal capabilities  &  Loss of control and potentially collision      & Malware injection to other ECUs via Bluetooth's ECU  & Vulnerability in \newline Bluetooth pairing mechanism &     ~\cite{Checkoway2011Comprehensive}  \\
		\hline
		Malware \newline injection 2	&    Telematics unit  cellular interface     &     Malware injection     &  Loss of control over other ECUs &  Remote control  and cyber-physical attacks    & Call car and inject malware payload  & Knowledge of car's specifics &     ~\cite{Checkoway2011Comprehensive}  \\
		\hline
		Malware \newline injection 3 & An ECU & ECU becomes an attack vector to inject undesired packets & CAN bus traffic becomes vulnerable to malicious packets & Other ECU behaves undesirably affecting cyber-physical components & Transmit malware in CAN packets & Physical access or vulnerable wireless interfaces & ~\cite{Experimental2010Koscher,Hoppe2011Security} \\
		\hline
		Packets injection  & Varies, depending on target ECU & Varies & Other ECUs and physical components & False data injection, loss of control, DoS, and safety-critical consequences & Compromised ECUs  inject packets & Malware injection & ~\cite{Experimental2010Koscher,Hoppe2011Security}                     \\
     	\hline
		Replay Attack & Lights ECUs & Lights turned off & Driver, passengers, and surrounding cars & Safety-critical situation & Eavesdrop and retransmit legitimate commands & Access to CAN bus network &     ~\cite{Experimental2010Koscher}  \\
       \hline
		Car's spying      & TCU    & Gain control of car  & All other ECUs can be remotely controlled &  Stealthily turn on car's microphone   & Call the car & Buffer overflow vulnerability and flaw in authentication protocol    & ~\cite{Checkoway2011Comprehensive}                       \\
		\hline
	    Relay attack & RKE system  & Open, and start a car without owner's knowledge & Target car& Theft and unauthorized access & Capture and relay LF beacon signals from car to key fob, and relay resulting UHF signal from key fob to car & Attacker needs relaying tools e.g., antennas and amplifiers among other devices & ~\cite{Francillon2011Relay}                       \\
       \hline
	\end{tabular}
	\caption{Smart Cars  Cyber-Physical attacks}
	\label{CarsAttacks}
		\end{footnotesize}
\end{table*} 

\section{Security Controls/Solutions}\label{sec:controls}

We briefly describe the research trends in CPS controls in two distinct paths. The first one is the solution that targets CPS in general, regardless of the application. The second is application-specific solutions that are specifically designed for some applications. In addition, we will highlight, whenever applicable, some solutions that can be cross-domain. That is, for example, some solutions designed for cars could be applied to medical devices, or vice versa.

\subsection{General CPS Controls}

Here we review controls that consider securing CPS, regardless of the application.
Addressing the vulnerability causes is the first step in the solution.

\emph{Controls against more connectivity}.
New security considerations must be taken into account to secure the access point from unauthorized access. Furthermore, the communication protocols used for realizing such connectivity are either proprietary protocols, such as Modbus and DNP3 in deployed ICS and smart grids, or open protocols such as TCP/IP. The proprietary protocols are burdened with a lot of vulnerabilities due to the isolation assumption when the protocols were designed~\cite{Amin2013IEEENet}.

\emph{Communication controls}.
Security solutions at the communication level in ICS should consider the differences with traditional IT solutions. For example, Intrusion Detection Systems (IDS) should be time-critical so that long delays are intolerable~\cite{Mitchell2013Survey}. In ~\cite{Mitchell2011IWCMC,Mitchell2013Effect}, the authors focus on designing IDS solutions for CPS. They also provide a comprehensive survey on IDS solutions in CPS applications~\cite{Mitchell2013Survey}.

\emph{Device Attestation}.
CPS components running software need to verify the software's authenticity. This verification helps significantly minimize malware. For example, hardware-based solutions such as Trusted Platform Module (TPM) provides code attestation. However, TPMs are assumed to be physical secure, which is infeasible to guarantee in some CPS applications such as ICS and smart grids. Another problem with TPM is the computational overhead on the limited resource CPS applications. Therefore, a new generation of TPMs that considers the limited CPS resources is needed.

\subsection{System-specific Controls}

\vspace{1mm}\noindent\textbf{ICS Controls.}

\emph{New design}.
ICS needs security solutions that are specifically designed for it. Such solutions should take into consideration the cyber-physical interactions, and the heterogeneity of components and protocols. Cardenas et al.~\cite{Cardenas2009Challenges} suggest that, most of the solutions in ICS aim to provide reliability, i.e., they make ICS reliable in the presence of non-malicious failures. Although reliability is important, malicious cyber attacks are now possible more than ever, and must be considered when designing new solutions. Therefore, security is as important as reliability.


\emph{Protocols with add-on security}.
Security solutions at the ICS communication level should consider the fundamental differences from traditional IT solutions. A number of proposals that rely on modifications of currents protocols, such as Modbus, DNP, and ICCP, to integrate security. Fovino et al.~\cite{fovino2009design} proposed the \textit{Secure Modbus} framework. It provides authentication, non-repudiation, and thwart replayed packets. Majdalawieh et al.~\cite{Majdalawieh2006Dnpsec} proposed \textit{DNPSec} framework, which adds confidentiality, integrity, and authenticity.

\emph{IDS}.
The complexity of designing IDS for securing ICS is relatively less than it is in traditional IT security. This is due to the predictability of the traffic and the static topology of the network~\cite{Krotofil2013INDIN}. Zhu et al.~\cite{Zhu2010Scada} defined a set of goals that IDS in ICS are expected to monitor such as detection of 1)any access to controllers and sensors/actuators' communication links, 2)modifications in sensor settings, and 3)actuators' physical tampering. D`Amico et al.~\cite{DAmico2011Integrating} propose \textit{WildCAT}, a solution that targets the physical exposure of the wireless communications within an ICS plant. It is a prototype for securing ICS from cyber attacks that exploits wireless networks. The idea is to install WildCAT in security guards' cars and to collect wireless activities in the physical perimeter of the plant. The collected data is sent to an analysis center, which, in turn, detects any suspicious activities and direct the guards to the location causing such activities. For further analysis of the current ICS-specific IDS solutions, we refer you to~\cite{Zhu2010Scada,Briesemeister2010Detection,Mitchell2013Survey,Krotofil2013INDIN,Cheminod2013Review,barbosa2014anomaly,Krotofil2015Process}.

\emph{Remote access to field devices}.
Fernandez et al.~\cite{Fernandez2005Scada} suggest that only authorized personnel can remotely access field devices. In addition, the access should be strictly secured by using a designated laptop through a VPN. In addition, Turk et al.~\cite{turk2005cyber} suggests a simple control for web-based DoS attacks that field devices with web access features are vulnerable to. The author suggests to close idle connections. In addition, it could be a good measure to disallow multiple connections simultaneously. Usually, no more than one legitimate employee tries to access such a resource at the same time.

\emph{Encryption and key management}.
There is undoubtedly a need for encryption in ICS networks. One of the associated problems with encryption is the delay, which is not desirable in time-sensitive environments. Choi et al.~\cite{Choi2009Advanced} proposed an ICS-specific key management solution that does not cause delay. In addition, Cao2013ALayeredet al.~\cite{Cao2013ALayered} proposed a layered approach aiming to protect sensitive data in the widespread ICS environments. Their technique relies on Hash Chains to provide: 1) layered protection such that ICS is split into two zones: high and low security levels, and 2) a lightweight key management mechanism. Thanks to the layered approach, an attacker with full control of a device in the low security level cannot intercept data from higher security level zones.

\emph{Software controls}.
Regular patching of security vulnerabilities in operating systems and their applications is a vital security practice. Windows released Stuxnet-related security patches, without which, Stuxnet would have still been present~\cite{Langner2011Stuxnet}. However, vendors of ICS applications must also keep up with the patching and release compatible versions of their applications. This ensures that ICS operators do not resort to older versions of vulnerable OS to be able to use the compatible ICS application~\cite{Johnson2010Survey}.

\emph{Standardization}.
The National Institute of Standards and Technology (NIST) is one of the leading bodies in the standardization realm. Following ICS security standards like theirs, among others, should significantly contribute to securing ICS. For example, Stouffer et al.~\cite{Stouffer2011NIST82} provided a comprehensive guidelines for ICS security. They provide guidelines for technical controls such as firewalls, IDS, and access control, and operational controls such as personnel security, awareness, and training. In fact, technical and operational controls must always go hand in hand, and the negligence of one leads to serious attacks. For example, lack of awareness could make employees vulnerable to social engineering attacks such as phishing. ICS-CERT reported that most of the attacks on ICS originated from phishing emails with malware-infected attachments~\cite{alcaraz2013critical}. In addition, security experts evaluated the security of an ICS corporation, and were able, through social engineering and phishing, to gain employees' credentials~\cite{Nicholson2012ComandSec}. Sommestad et al.~\cite{Sommestad2010Scada} conducted a comparison, based on keywords mining, and concluded that the standards focus either on technical controls, or operational controls, but not both. In addition, some standards somewhat neglect ICS-specific properties and focus on IT security countermeasures alone.

\vspace{2mm}\noindent\textbf{Smart Grids Controls.}

\emph{DoS controls}.
Attacks on communication components should be prevented or, at least, detected. On the one hand, at the network layer, prevention of attacks like DoS is usually achieved by rate-limiting, filtering malicious packets, and reconfiguration of network architecture. The first two are possible in smart grids, while the last might be difficult due to its relatively static nature. Furthermore, techniques at the physical layer aim at preventing attacks of the nature of wireless jamming. On the other hand, DoS detection techniques are categorized into four types: signal-based, packet-based, proactive, and hybrid detection~\cite{Wang2013Cyber}.


\emph{IDS}.
IDS for smart grids is still an ongoing problem that is not that mature yet. Designing IDS for smart grids is a complex task due to the enormous size of the grids and the heterogeneity of their components~\cite{Sridhar2012IEEE}. In addition, IDS built for traditional IT systems will not necessarily work for the smart grids. They must be specifically designed for smart grids to reduce the likelihood of false detection rates. Jin et al.~\cite{Jin2006Anomaly} proposed an anomaly-based IDS that detect malicious behavior using invariant detection and artificial ants with Bayesian reasoning approach. In addition, Mitchell and Chen~\cite{Mitchell2013Behavior} proposed a behavior-rule-based IDS to detect attacks on cyber-physical devices in smart grids such as headends, subscriber energy meters (SEMs), and data aggregation points (DAPs). Liu et al.~\cite{Liu2015Abnormal} proposed an IDS for detecting bad data injection attacks targeting the smart grids. Their approach relies on combining detecting techniques from the traditional IDS and physical models.

\emph{Low-level authorization and authentication}.
 A common problem in a large system like smart grids is authentication and authorization of users who need to gain access to low-level layers such as field devices. Commonly, all field devices share the same password that employees know. This results in the impossibility of the non-repudiation security requirement. A malicious employee could gain access to a field device and make undesired changes to the system, and there is no way to trace who did it. Therefore, Vaidya et al.~\cite{Vaidya2013Authentication} proposed an authentication and authorization mechanism that provides legitimate employees the ability to access field devices in the substation automation systems in smart grids. Their proposal relies on elliptic curve cryptography due to its low computation and key size requirements compared with other public key mechanisms.

\emph{New designs}.
New security issues require that various aspects of smart grids be approached differently. The cyber-physical nature of the systems needs to be considered. Mo et al. ~\cite{Mo2012IEEE} proposed \textit{Cyber-Phyiscal Security}, a new approach that combines systems-theoretic and cyber security controls. They provide two examples showing the applicability of their approach on two attacks on smart grids: replay attack, and stealthy deception. They emphasize the need for considering those two types of components, cyber- and physical-components, when designing controls for smart grids. Most of the work done is extending existing protocols and systems to capture security properties. This might work as a temporary solution, but a bottom-up redesign is desired.

\emph{Security extensions}.
The trend of adding-on security to existing components of smart grids has been emerging. Protocols like DNP3, IEC 61850 and IEC 62351 are extended to capture security properties. For example, Secure DNP3 protocol is an extended DNP3 that have basic authentication, integrity and confidentiality services. The security features are added by inserting a security layer in the communication stacks of these protocols~\cite{Wang2013Cyber}.

\emph{Privacy-preserving controls}.
 Lack of confidentiality in data aggregation protocols might result in privacy invasion of consumers' private information such as billing information and usage patterns~\cite{McDaniel2009SPIEEE}, while the lack of integrity could result in disruption in state estimation and consumption reports~\cite{Sridhar2012IEEE}. Therefore, a number of privacy-preserving techniques have emerged to provide aggregated data with confidentiality and integrity when in transit between smart meters and control centers ~\cite{Wang2013Cyber,Fang2012Smart}. Another privacy concern that might affect safety or finance is the ability to detect the (in)occupancy of house in order to break in. Chen et al.~\cite{chen2015preventing} proposed \textit{combined heat and privacy} (CHPr) mechanism such that it makes the poser usage data always looks like the house is occupied and, therefore, tricks occupancy detection techniques.





\emph{Standardization}.
A number of bodies, such as the IEC and NIST, have developed a set of standards for securing smart grids' communications. For example, IEC's have developed standards TC57 and 6235~\cite{Cleveland2012IEC}, while NIST has developed guidelines for smart grids in report 7628~\cite{Nistir20107628}.

%

\emph{Smart meters' disabling prevention}.
To prevent remote attackers who exploit the disabling feature in smart meters, Anderson and Fuloria~\cite{Anderson2010Controls} suggest that smart meters should be programmed to notify customers in enough time in advance, before the command takes effect and disables meters. This measure helps in the early detection of DoS attempts before they take place.

\emph{Physical security}.
As smart meters are physically exposed, they must be physically protected. NIST standards~\cite{Nistir20107628} state that smart meters must have cryptographic modules in addition to physical protection. The standards also emphasize on the need for smart meters to be sealed in tamper-resistant units such that unauthorized parties are not able to physically tamper with them.

\vspace{2mm}\noindent\textbf{Medical Devices Controls.}

\emph{Authentication}.Halperin et al.~\cite{Halperin2008Pacemakers} proposed a cryptographic-based authentication and key-exchange mechanism to prevent unauthorized parties from accessing IMDs. Both mechanisms do not consume batteries as a source of energy. Instead, they rely on external radio frequency. In addition, Out-of-Band (OBB) authentication is deployed in some wearable and implantable devices. By which, authentication is performed using additional channels, other than the channels used for communication, such as audio and visual channels~\cite{Rushanansok2014SoK}. In addition, biometrics, such as electrocardiograms, physiological values (PVs), heart rate~\cite{seepers2016secure}, glucose level and blood pressure, can all be used for key generation for encrypted communication in the body sensor network (BSN)~\cite{Rushanansok2014SoK}. In addition, patients' movement can be another property by which keys are derived~\cite{oberoi2016wearable}.

\emph{Intrusion Detection Systems}.
Halperinet al.~\cite{Halperin2008Pacemakers} proposed a detection mechanism that alarms patients of unauthorized communication attempts with their IMDs. In addition, Gollakota et al.~\cite{Gollakota2011They} proposed the \textit{Shield}, which detects and prevents malicious wireless-based attacks on IMDs. Although the \textit{Shield} is not designed specifically as an IDS, it certainly serves as one. On the other hand, Mitchell and Chen~\cite{Mitchell2012Behavior} aim to detect compromised sensors and actuators that pose threat to patients' safety through behavior rule-based IDS. Their proposal is not intended for IMDs or wearable devices. Rather, it is mainly for stand alone medical devices, such as \textit{vital sign monitor} and \textit{cardiac device}. Thus, there is a need for IDS solutions that consider implanted and wearable devices' unique properties, e.g., communication protocols, physical interaction with human bodies, and limited resources.


\emph{Location-based controls}. Some solutions utilize \textit{distance-bounding} protocols that rely on the physical distance between communicating devices so that remote attackers cannot launch attacks remotely. The distance is determined by various techniques such as ultra sound signals, received signal strength (RSS), electrocardiography (ECG) signals~\cite{zheng2015encryption}, and body-coupled communication (BCC). This technique provides authentication but not authorization. Hence, other techniques must be incorporated~\cite{Rushanansok2014SoK}.

\emph{Thwarting active and passive attacks}.
Li et al.~\cite{Li2011Hijacking} proposed the use of BCC, were they experimentally investigate the BCC's ability to prevent passive and active attacks against insulin delivery systems. This type of communication thwarts most passive and active attacks because of its dependence on the human body as its transmission medium, as opposed to conventional wireless communication where the air is the communication medium that is easily intercepted.  When the human body becomes the transmission medium, an attacker needs a very close proximity to a patient or even direct body contact. This significantly mitigates the attacks and raises the bar for the attackers.

\vspace{2mm}\indent\emph{Shifting security to wearable devices}.
 Incorporating security into the current IMDs and wearable devices has its own risks and challenges. One of which is the health risk associated with IMDs' surgical extraction from a patient in order to update or replace an IMD with more secure one. In fact, even if we assume that there are no health risks for extracting IMDs, the cryptographic operations required for any secure system are still expensive in terms of computational, memory, and battery resources. Therefore, the intuitive solution is to add another device built specifically to add security. Several proposals that deploy some cryptographic and anti-jamming-attack mechanisms utilize external wearable devices to realize such mechanisms. For example, Xu et al.~\cite{Xu2011IMDGuard} proposed \textit{IMDGuard} to defend against jamming and spoofing attacks. In addition, Gollakota et al.~\cite{Gollakota2011They} proposed an external wearable device, the \textit{shield}, to detect and prevent any unauthorized commands sent to an IMD. They evaluated the \textit{shield} on two modern IMDs, i.e., ICD and cardiac resynchronization therapy device (CRT). This device jams any signals initiated by unauthorized party.


\emph{Cross-domain solutions}.
Li et al.~\cite{Li2011Hijacking} proposed the adoption of the \textit{rolling code} encryption mechanism used in RKE in cars. Smart cars and medical devices both share similar features in terms of computation limitations, power, and data bandwidth constraints. Therefore, using rolling code encryption should be an effective solution to prevent eavesdropping and replay attacks.

\emph{Standardization and recommendations}.
The Food and Drug Administration (FDA) is the leading body in medical devices' standardization. It has issued a number of standards and guidelines for the manufacturers of medical devices. For example, in 2005, the FDA highlighted that potential vulnerabilities might result from using COTS software equipped with remote access capabilities~\cite{FDA2005Cybersecurity}. Another recommendation was published in 2014 about cybersecurity in medical devices~\cite{FDA2014Cybersecurity}. However, the recommendations are not detailed enough nor mandatory, rather ''non-binding recommendations". Therefore, manufacturers have the liberty to choose not to follow them, which certainly would contribute to the production of less secure medical devices. In addition, IEEE 802.15.6 is the latest BAN standard that provides security services such as authentication and encryption~\cite{IEEE2012BAN}

\emph{Allowing vs. disallowing remote functionalities}.
In order to prevent attackers from penetrating networks that make the interaction between remote physicians with patients' devices possible, manufacturers should disable remote capabilities from being sent through the network. They only allow remote parties to receive measures and logs, but not send commands. Although this is a good security practice to prevent attackers from sending remote commands, it limits the full utilization of such devices~\cite{Lee2012Challenges}. Therefore, there is a need to strike a balance between security and usability without introducing remote threats to patients. If remote commands are allowed, Hayajneh et al. ~\cite{hayajneh2016secure} proposed an approach to protect patients from unauthenticated remote commands against their IMDs. The approach relies on Rabin public-key cryptosystem.

%

\vspace{2mm}\noindent\textbf{Smart Cars Controls}

\emph{Unimplemented promising controls}.
A number of security controls have been proposed to secure the in-car network, most of which have not been implemented. For example, Wolf et at.~\cite{Wolf2004Securityin} proposed three controls that would secure the bus network: authentication gateway, encryption, and firewalls. In addition, Larson et al.~\cite{Larson2008Securing} call for redesigning security in cars, and propose embracing of the \textit{defense-in-depth} security paradigm, i.e., prevention, detection, deflection, countermeasures, and recovery.

\emph{Cryptography}.
The use of cryptography provides a number of security properties such confidentiality, integrity, and authentication that cars lack. However, these mechanisms are computationally expensive for such systems with limited capabilities in cars. Thus, the deployment of efficient solutions is vital. Wolf et al.~\cite{Wolf2012Design} and Escherich et al.~\cite{Escherich2009She} propose hardware-based solutions that are designed specifically for cars' security. Wolf et al.~\cite{Wolf2012Design} designed and implemented the Hardware Security Module (HSM). They show its applicability to secure communications of ECUs within a car, or even in Vehicle-to-Vehicle (V2V) communications. Escherich et al.~\cite{Escherich2009She} presented the Secure Hardware Extension (SHE), a standard for adding security properties, such as secret key protection and secure boot, to ECUs.

\emph{Redefining trust}.
Koscher et al.~\cite{Experimental2010Koscher} suggest two trust-related controls that would have prevented most, if not all, of their attacks. Firstly, revoking trust from arbitrary ECUs so they cannot perform diagnostic and reflashing operations. Secondly, ECUs with diagnostic and reflashing capabilities must be authorized and authenticated before performing these tasks. To do that, trusted platforms, in addition to remote attestation, need to be deployed~\cite{Kleidermacher2012Embedded}.


\emph{Restricted critical commands}.
Koscher et al.~\cite{Experimental2010Koscher} emphasized that legitimate parties require physical access to cars before  issuing any ``dangerous" commands. Although this could be an effective control, the term dangerous is relative, and its interpretation varies from one manufacturer to another, so that seemingly-benign commands could result in serious attacks. If manufacturers decide to restrict the amount of commands that require physical access, flexibility and convenience will be affected. Therefore, we need solutions that consider all possible attacks resulting from critical or benign commands, while maintaining the existing flexibility.

\emph{Bluetooth}.
Bluetooth connections between devices and cars can be exploited to launch different attacks such as compromising the TCU and consequently other ECUs~\cite{Checkoway2011Comprehensive}. Cars need an additional security layer to defend against Bluetooth-dependent attacks. Dardanelli et al.~\cite{Dardanelli2013Security} show the applicability of their proposed security layer to protect against smartphone-initiated Bluetooth attacks, with little impact on performance. Although their proposal was tested on a two-wheeled vehicle, it should also be applicable to cars. Furthermore, Woo et al.~\cite{woo2015practical} proposed a security mechanism to efficiently authenticate connecting devices to smart cars in order to prevent wireless attacks exploiting vulnerabilities in smartphones.

\emph{IDS}.
Most of the proposed IDS are designed for CAN protocol, and only a few for other protocols such as FlexRay and LIN. For example, Larson et al.~\cite{Larson2008Approach} proposed a specification-based IDS that is implemented in each ECU, whereas ~\cite{Seifert2014Secure} proposed a behavior-based IDS that supports FlexRay networks, besides CAN. In addition, Miller and Valasek~\cite{Miller2014aSurvey} demonstrated a proof-of-concept low-cost attack detection system that detects anomalies in the CAN network, and the great opportunities for implementing such a system at low cost and no manufacturing overhead. Furthermore, Cho and Shin~\cite{cho2016fingerprinting} proposed an anomaly-based IDS that is clock-based such that it measures and utilizes the intervals of periodic messages in order to uniquely identify ECUs. The authors call it ``fingerprinting". Taylor et al.~\cite{taylor2015frequency} proposed a frequency-based IDS that detects anomalies between the frequency of current and historical packets that have strict frequencies.

\emph{Physical Attacks Controls}.
Shoukry et al.~\cite{Shoukry2015Pycra} proposed a challenge-response authentication scheme that works at the physical level. The scheme detects and prevents physical attacks such as sensors spoofing. The scheme utilizes the physical properties of signals in the physical/analog domain by which they were able to implement their scheme. The scheme detects and prevents the demonstrated attack on the ABS~\cite{Shoukry2013Non}.

\subsection{Cyber-Physical Security Framework}

Now after we have surveyed security in the four CPS applications, we shed light on the relationship between the CPS and security aspects. In Table~\ref{Framework}, we summarize the four security issues: (1) \emph{threats}, (2) \emph{vulnerabilities}, (3) \emph{attacks}, and (4) \emph{controls} with respect to three CPS aspects: (1) \emph{cyber}, (2) \emph{cyber-physical}, and (3) \emph{physical} for ICS, smart grids, medical devices, and smart cars. Please note that the table represents a sample of the surveyed aspects under our framework, and is by no means comprehensive.


\begin{table*}
	\begin{footnotesize}
	\begin{tabular}{p{2cm}p{2cm}p{4cm}p{4cm}p{4cm}}
		\toprule
\textbf{CPS} & \textbf{Threats} & \textbf{Vulnerabilities} & \textbf{Attacks} & \textbf{Controls} \\ \midrule
		ICS    & Criminal (C)      &  Internet-connected field devices (C) & DoS attacks (CP) & Close timed out connections (C) \cite{turk2005cyber}  \\
		    & Criminal (C)      &  COTS TCP/IP stack flaw (C) & Malware injection that restores field devices to default (CP) & TCP/IP supplier fixes the flaw (C) \cite{turk2005cyber}  \\
		    & Espionage (C)     & Personnel lack of awareness (C) & Dragonfly attack (C) & Raise personnel awareness of attack entry points and use proper antivirus and IDS mechanisms (C) \cite{Symantec2014Dragonfly} \\
		    & Cyberwar  (C)    &    &      Stuxnet: damaged about 1000 centrifuges (CP) \cite{W322011Falliere} & Raise personnel awareness of the impact using infected devices/media (CP), regular software patching (C) and encryption (C) \cite{Choi2009Advanced}  \\
		    & Insider (CP)      &   Reliance on security by obscurity \& Lack of access control and non-repudiation (C) & Maroochy: a ex-insider exploits knowledge of a systems' weakness \cite{Slay2007Lessons} & Fine-grained access control (C \& CP) and security by design (C)~\cite{Stouffer2011NIST82}  \\
     		& Customer (CP)      & Physical exposure of ICS components (P) & Sensors tampering to cause false reading (CP)  & Tamper-resistant field components (P)\cite{Krotofil2013INDIN}  \\ \midrule
		Smart Grid    & Criminal (C)      &       Internet-connected components of the grid (C)&  Take control and cause blackouts (CP) \cite{Nakashima2008Hackers} &  Avoid default and simple passwords (C) \& Security-in-depth \\
		   & Hostile nations \& criminals (C)      & Aurora vulnerability: power generators' breaker & Disruption of circuit breakers to damage generators & Breakers' protection measures \cite{Zeller2011Myth}  \\
		   & Espionage (C)      &  Unencrypted or inferable customers' data (C) &   Infer data and reveal customers' private information, habits, \& appliances (C) & Encryption (C) \cite{Li2010Secure,Yan2012ASurvey}  \\
		   & Customer (CP)      &  Physical exposure of smart meters (P)  & Physical tampering with smart meters to reduce utility bills\cite{Komninos2014Survey} & Tamper-resistant meters \cite{Cho2014Privacy}  \\
		   & Thieves \& vandals (P)      &     Unprotected field devices and other equipments and cables (P) & Stealing or damaging such components affects the grid and its customers (CP) & Physical protection and improved resiliency (P) \cite{Cleveland2012IEC} \\
		   \midrule
		Medical Devices    & Criminal (C)      &  Insecure wireless transmission between programmers and ICDs (C) & Turing off ICDs by retransmitting ''turn off" command (CP) & Zero-power detection \& prevention mechanism (C) \cite{Halperin2008Pacemakers}   \\
		   & Criminal (C)  &   Insecure wireless communications between insulin pump, its remote control, and CGM (C) & FDI (C), command injection (CP), \& replay attacks (CP) & Encrypted communication based on rolling-code \& body-coupled  communication \cite{Li2011Hijacking} \\
		   & Espionage (C)      &  Weak or no encryption (C) &  Revealing patients' private information (C) \cite{Halperin2008Pacemakers,Li2010Secure} & Use lightweight encryption (C)\cite{Halperin2008Pacemakers} or wearable security devices (CP)\cite{Gollakota2011They,Xu2011IMDGuard,Denning2008Absence}   \\
		   & Politically-motivated  (C)    &  Wireless control capability (C) & Target a political figure by tampering with ICD configurations (CP) \cite{Rushanansok2014SoK} & Zero power encryption mechanism (C) \cite{Halperin2008Pacemakers} or wearable devices (CP) \cite{Denning2008Absence,Gollakota2011They,Xu2011IMDGuard}\\
		   \midrule
		Smart Cars    & Thieves (P)    &  Tampering (P) \cite{Hoppe2011Security,Wolf2004Securityin} &  FDI (C) \cite{Experimental2010Koscher,Hoppe2011Security} &  Hardware-based security measures \cite{Escherich2009She,Wolf2012Design} \\
				      & Criminal (C)      &   RKE system (P) &  Relay signals between car owner's key and car (P) \cite{Francillon2011Relay} & Key shielding (P) and distance-bounding countermeasures (CP) \cite{Francillon2011Relay} \\
		              & Espionage (C)      &  Wireless links (C) in TPMS\cite{ishtiaq2010security}, cellular\cite{Miller2014aSurvey,Checkoway2011Comprehensive}, \& Bluetooth\cite{Checkoway2011Comprehensive}  &  Privacy invasion (C)\cite{Checkoway2011Comprehensive}       & Secure Bluetooth (C)\cite{Dardanelli2013Security} \& trust redefinition of COTS products (C)\cite{Experimental2010Koscher}\\
		              & Criminal (C) & Vulnerability in media player's parser (C) \cite{Checkoway2011Comprehensive}& DoS and lost of control (CP) \cite{Checkoway2011Comprehensive} & Trust redefinition of COTS products (C)\cite{Experimental2010Koscher}\\
	 \bottomrule
	\end{tabular}
	\caption{Cyber-Physical Security Framework. C: Cyber, CP: Cyber-Physical, and P: Physical}
	\label{Framework}
	\end{footnotesize}
\end{table*}

%

\section{CPS Security Challenges}\label{sec:challenges}
\vspace{2mm}\noindent\textbf{General CPS Security Challenges}

\emph{Security by design}. Security is not taken into consideration in the design of most CPS as a result of their isolation in physically-secured environments without connectivity to other networks, the Internet for instance. Hence, physical security has been almost the main security measure~\cite{Stouffer2011NIST82}.

\emph{Cyber-physical security}.
 The security mindset of CPS designers needs to change so that they consider both cyber and physical aspects. This way potential cyber-attacks with physical consequences will be better predicted and thus mitigated~\cite{Gollmann2013Security}. Neuman et al.~\cite{Neuman2009Challenges} suggests that when the fundamental differences between cyber and physical aspects are not considered, cyber-physical solutions are usually ignored, and the focus becomes cyber-only solutions. This urges the need for considering both cyber- and physical-aspects. Mo et al.~\cite{Mo2012IEEE} described a new field called ``cyber-physical security". The authors' aim to help by their proposal in developing novel solutions for the CPS security issues, especially in the context of smart grids. Furthermore, the survivability of the systems under attack is very crucial in CPS. A set of CPS security challenges, such as survivability, are discussed in~\cite{Cardenas2009Challenges}.

\emph{The real-timeliness nature}.
The real-time requirement is a requirement whose absence affects the security posture~\cite{Cardenas2011Attacks,Neuman2009Challenges}. During a security attack, real-time decisions in CPS are crucial for systems' survivability. Therefore, consideration of the interactions between physical- and cyber-aspects in any CPS security design gives the full picture of the system, which assists in designing better risk-assessment, attack-detection, and attack-resilient solutions~\cite{Cardenas2011Attacks}. In addition, cryptographic mechanisms could cause delays that could affect some real-time deadlines. Therefore, lightweight and hardware-based mechanisms should be considered.

\emph{Uncoordinated Change}.
The number of CPS stakeholders is relatively large. This includes manufacturers, implementors, operators, administrators and consumers. Their activities and privileges differ, and hence need to be properly managed~\cite{Amin2013IEEENet}. The large number of stakeholders, as well as the heterogeneous CPS components, require \emph{change management}. This is another challenge that we observe to be somewhat ignored. When changes occur in a group of CPS components, some coordination is required at some level by the stakeholders. Examples of such changes are replacing hardware, updating or changing software, and adding new capabilities~\cite{Luallen2011Sans}. Any uncoordinated change might alter the initial assumptions about the CPS security, and therefore could introduce new vulnerabilities.

In addition to the aforementioned general challenges to CPS security, we highlight some application-specific security challenges in the following paragraphs.

\vspace{2mm}\noindent\textbf{ICS Challenges}

\emph{Change management}. The ICS environment spans diverse geographical locations that involve various systems that need to be replaced, updated, or removed at some point. For example, in ICS, a system update needs careful planning to avoid unexpected failure as it has occurred in a nuclear plant due to an update in a computer in the business network of the plant~\cite{Cardenas2009Challenges}. In addition, there is a large number of stakeholders who can affect the security posture unintentionally. Therefore, some kind of coordinated change management should be introduced to prevent and detect security-related changes in the ICS application~\cite{Stouffer2011NIST82,Luallen2011Sans}.

\emph{Insider threat}. This is probably one of the most difficult threats to defend against. Insiders could cause security problems either intentionally or unintentionally. For example, an insider could leverage the trust given to her/him along with acquired inside knowledge to launch an attack, such as the Maroochy water and sewage system, or help remote attackers, such as Stuxnet propagation via a USB stick. Unintentional cases where insiders unknowingly use an infected laptop or USB stick that could give remote attackers access points to the ICS. The insider threat has been underestimated and overlooked, and it certainly needs serious considerations~\cite{Krotofil2013INDIN}.

\emph{Secure integration}. As ICS relies heavily on legacy systems, it is inherently vulnerable to their vulnerabilities. Therefore, the integration of new components with legacy systems must be done securely so that it does not result in new security vulnerabilities. In addition, due to the large number of legacy components in ICS, it is economically infeasible to replace all of them with more secure ones~\cite{Krotofil2013INDIN}. Meanwhile, their security should not be ignored and short-term solutions must be implemented to minimize any potential risks~\cite{Cardenas2009Challenges}.

\vspace{2mm}\noindent\textbf{Smart Grids Challenges}

\emph{Two-way communication}. One of the distinguishing features of smart grids is the two-way communication, thanks to the advanced metering infrastructure (AMI). Unlike the power grid, AMI allows smart meters attached to consumers' houses, that are easily accessible by physical attackers, to communicate with utility companies. This raises a new challenge to secure these devices~\cite{Khurana2010SPIEEE}.

\emph{Access control mechanisms}. Due to smart grids' enormous geographical coverage, in addition to the large number of stakeholders, appropriate access control mechanisms are needed~\cite{Amin2013IEEENet}. Every possible access to smart grids' network, data, or devices must be controlled and managed. In addition, during emergency, access control mechanisms need to have enough flexibility to give appropriate privileges for the appropriate parties.

\emph{Privacy concerns}. As consumers' data has become a significant part of the smart grids' traffic, privacy concerns have become a great challenge. Not only should consumers' data be encrypted, but anonymization techniques are also desired to prevent inference and other attacks from deducing patterns from the encrypted data to reveal private information~\cite{Komninos2014Survey,Molina2010Private}. Some cryptographic-based solutions have been proposed. Li et al.~\cite{Li2010Secure} proposed a homomorphic encryption mechanism to protect consumers privacy while maintaining low overhead on smart grids' traffic. However, such an approach does not prevent an attacker from participating in the data aggregation as a smart meter by injecting false data or impersonating a legitimate smart meter~\cite{Wang2013Cyber}. Therefore, designing mechanisms that both encrypt and aggregate data securely is a pressing challenge~\cite{Wang2013Cyber}.

\emph{Explicit trust}. Sensed data and sent commands should not be explicitly trusted. Instead, new mechanisms are needed to detect false data and unauthorized commands~\cite{Das2012Handbook}. With the large size of smart grids, it becomes difficult to detect false data injection attacks by relying on algorithms that have been designed to only detect faults~\cite{Liu2011False}.

\emph{Comprehensive security}. Security measures and tools mainly exist at higher levels in smart grids and their effectiveness decreases towards lower levels. In other words, the sophistication of security measures decreases in lower levels due to the limited capabilities in low-level devices. Hence, security needs to be involved in every part of smart grids, starting from the lowest levels, i.e., field devices and their protocols, to high levels, e.g., control centers. Implementing security at lower levels might have some performance costs. Therefore, lightweight solutions are desired~\cite{Knapp2013Applied}. The use of encryption is necessary to provide confidentiality and integrity at all levels of smart grids. However, the challenge is not in deploying it, rather, it is in doing it cost-effectively in low level components.

\emph{Change management}. Managing changes in smart grids is no less challenging than it is in ICS. Smart grids are certainly more diverse and have more stakeholders than ICS applications, and yet its change management capabilities are limited~\cite{Sridhar2012IEEE}. This makes change management an immensely desired requirement for a more secure smart grids.

\vspace{2mm}\noindent\textbf{Medical Devices Challenges}

\emph{Security vs. Usability}.
Too much security could be result in counterintuitively result in the inability to reconfigure a device when a patient is in a critical condition. For example, a patient with an IMD might be in a situation that needs urgent intervention by another health care provider. The provider does not have the cryptographic credentials or the access privileges that allow him/her to reconfigure the IMD, so the unavailability of the IMD could be very dangerous~\cite{Halperin2008Security,Rostami2013Balancing}. Therefore, designers should strike a balance between usability and security in medical devices. The usability property should allow access in emergency situations, for example, while maintaining security as much as possible. Denning et al.~\cite{Denning2014Cps} proposed a promising solution which uses a \emph{fail-open/safety wristband}. A patient wears the band to prevent interactions with unauthorized programmers and other illegitimate parties. Whenever there is a need to access an IMD, this band is removed, allowing communication with any programmer. The authors also propose a number of secure design considerations for maintaining usability and security in.

\emph{Add-on security $\approx$ increased code}. Adding security directly to the IMDs could increase the size of code and hence the rate of medical device recalls. In addition, cryptographic operations could also affect the limited-power at the cost of the original purpose of a medical device~\cite{Rushanansok2014SoK}. Thus, it is more desirable to limit functions of IMDs to pure medical operations and shift security services to an external device~\cite{Gollakota2011They}.

\emph{Limited resources}.
Extensive computations that are usually needed in cryptographic mechanisms consume power, which is a critical resource in these small and limited-resource devices~\cite{Halperin2008Security,Rostami2013Balancing}. The devices must maintain power for a number of years depending on the nature of the device. For devices that require surgery to be placed into a patient's body, they must be able to function for at least a decade or two. Halperin et al.~\cite{Halperin2008Pacemakers} proposed one of the first efforts to combine cryptographic techniques with battery-free consumption. The authors designed a cryptographic-based solution that relies solely on RF as a source of energy.

In addition, some attacks might only aim to drain the battery to disable a device, resulting in a DoS attack~\cite{Rushanansok2014SoK}. Although a medical device might refuse to interact with an unauthorized party sending signals, the very fact that the device receives and processes the signals from illegitimate parties is problematic and results in battery depleting. Therefore, new controls should be developed to prevent such attacks by preventing medical devices from responding to any illegitimate interaction.


\vspace{2mm}\noindent\textbf{Smart Cars Challenges}

\emph{Secure integration}.
Incompatible security assumptions occur at the boundaries of the integrations when manufacturers integrate COTS and third party components into smart cars. The lack of the products' internal details results in this mismatch. Therefore, manufacturers need to ensure a secure integration of COTS and third party components. Sagstetter et al.~\cite{Sagstetter2013Security} suggest the adoption of formal methods such as a model-based design that is combined with verification methods to verify the correctness of any assumptions about COTS and third party components. In addition, access control measures need to be implemented to prevent unauthorized operations, especially that originate from COTS and third party components~\cite{Nilsson2008Vehicle}. This ensures that the security posture assumed by a car's manufacturer is not compromised.

\emph{Effective separation}. Although gateway ECUs are supposed to separate CAN network traffic into high and low bandwidths, various attacks were able to bypass it and gain access to restricted bandwidths~\cite{Experimental2010Koscher}. However, some manufactures actually have deployed effective separation between critical and non-critical ECUs by deploying them in completely separate networks~\cite{Miller2014aSurvey}. However, this practice is not that common among manufacturers. Another promising solution is the use of Ethernet/IP communications and the replacement of gateway ECUs by Master-ECUs~\cite{Sagstetter2013Security}. One the one hand, this should provide more bandwidth, so cryptographic solutions can work without causing communication overhead, and much higher speeds. On the other hand, it requires manufacturers to replace the legacy ECUs and network architecture which is a very costly operation.

%

\emph{Heterogeneity of components}. It is difficult to suggest that all components produced by different Original Equipment Manufacturers (OEMs) should be replaced by components produced only by car manufacturers. This might be an impractical solution given the complexity and highly skilled specialty involved in designing different components that are integrated with cars. Instead, both parties must be in accord in terms of security requirements, assessment and testing. Manufacturers must incorporate security engineers from the early design phase~\cite{Kleidermacher2012Embedded}.


\emph{In-car communication}. The CAN network is inherently vulnerable due to the isolation assumption, and thus new protocols that assume existing potential malicious attackers are needed. The OVERSEE project~\cite{Groll2009Oversee} aims to design such protocols that would revolutionize the legacy CAN protocol and replace it with a highly secure and dependable communication platform. Other temporary solutions such as firewalls and IDS are also essential, and could be incorporated into existing cars either as a part of the gateway ECUs or as stand alone ECUs. A current project on a \textit{stateful IDS} that have a contextual awareness of a current situation~\cite{Studnia2013Security}. For example, it aims to detect the legitimacy of messages sent while a car is in a specific status: parked, driving, and many other contexts.

\emph{New vulnerabilities and attacks}. In the forthcoming era of autonomous cars, V2V, and V2I communications, new security issues are certainly going to arise. Such potential attacks are discussed in~\cite{petit2015potential}.

\section{Conclusion}

In the paper, we survey the literature on security and privacy of cyber-physical systems, with a special focus on four representative CPS applications: ICS, smart grids, medical devices, and smart cars. We present a taxonomy of threats, vulnerabilities, known attacks and existing controls. We also present a cyber-physical security framework that incorporates CPS aspects into the security aspects.
The framework captures how an attack of the physical domain of a CPS can result in unexpected consequences in the cyber domain and vice versa along with proposed solutions. Using our framework, effective controls can be developed to eliminate cyber-physical attacks. For example, we identified that the heterogeneity of CPS components contributes significantly to many attacks. Therefore, an effective solution should pay special attention when heterogeneous components interact.

The research in CPS security is active because of the frequently reported cyber-attacks. Although some defense mechanisms have been proposed/deployed, new and system-specific solutions are still expected in response to the newly identified threats and vulnerabilities. In this paper, we also highlight challenges and some missing pieces in CPS security research, and hope to stimulate more interests in the research community.

\addtocontents{toc}{\protect\vspace*{\baselineskip}}
\Urlmuskip=0mu plus 1mu\relax
\bibliographystyle{plain} 
\bibliography{refs} 



\end{document}